\documentclass[intlimits,twoside,a4paper]{article}

\usepackage[cp1251]{inputenc}
\usepackage[eqsecnum]{cmpj3}

\usepackage{multirow}
\usepackage{booktabs}

\usepackage{bm}

\issue{2026}{29}{2}{23801}
\doinumber{10.5488/CMP.29.23801}
\title[Thermodynamic stability and structural transitions in virus--host networks]%
{Thermodynamic stability and structural transitions in virus--host networks}
\author[A. Rovenchak, M. Husiev]{A. Rovenchak\orcid{0000-0002-0452-6873}\refaddr{label1,label2}\thanks{Corresponding author: \email{andrij.rovenchak@lnu.edu.ua}.}, 
        M. Husiev\orcid{0009-0009-3001-1097}\refaddr{label1}}
\addresses{
\addr{label1} Professor Ivan Vakarchuk Department for Theoretical Physics, Ivan Franko National University of Lviv, 12~Drahomanov St., 79005 Lviv, Ukraine
\addr{label2} SoftServe, Inc., 2d Sadova St., 79021 Lviv, Ukraine
}


\usepackage{ulem}

\date{Received 26 February 2026; revised 26 April 2026; accepted 05 May 2026; published 29 June 2026}

\begin{document}

\maketitle

\begin{abstract}
Understanding virus--host interactions is crucial for predicting the stability of networks under various perturbations.
In this study, we present an analysis of virus-related networks for several organisms (\textsl{Homo sapiens}, \textsl{Mus musculus}, \textsl{Gallus gallus}), encompassing directed and weighted connections.
We compute a range of network parameters, including topological characteristics and thermodynamic quantities derived from adjacency spectra, to gain insights into the structural robustness and dynamic behavior of the networks.
To assess stability, we model two distinct node removal scenarios: targeted elimination of the most influential nodes and random removal.
Our findings reveal transition-like behavior in spectral thermodynamic functions and characteristic changes in structural measures, contributing to evaluating the potential of a thermodynamic framework for studying virus--host networks and advancing a deeper understanding of their dynamics.
%
%
\printkeywords
%
\end{abstract}

\section{Introduction}


Viral infections pose a serious threat to human populations, agricultural crops, and ecosystems as a whole \cite{fu2011imitation}.
Understanding the mechanisms of viral interactions with hosts, as well as the structural and functional characteristics of these processes, is one of the key challenges in modern bioinformatics and virology \cite{damas2020broad}.
In this context, network analysis is gaining an increasing importance, as it enables the identification of patterns in the organization of virus--host interactions, the assessment of their resilience to disruptive influences, and the development of new strategies for combating infectious agents \cite{gulbahce2012viral,vidal2011interactome}.

Viral interactions can be represented as complex networks \cite{guirimand2015virhostnet}, in which nodes correspond to viruses, proteins, RNA molecules, or other host cell components, and edges represent functional relationships between them.
These networks exhibit a high degree of topological heterogeneity, characterized by significant variability in node connectivity and distinct distribution patterns.
One of the key properties of such networks is their robustness to random node removal, which likely reflects biological mechanisms that ensure the resilience of viral populations \cite{avs2022virus,bosl2019common}.

At the same time, incorporating directed and weighted edges into the analysis significantly enhances the ability to study the dynamics of virus--host interactions \cite{fendt2022overview}, but also complicates the interpretation of the results.
Investigating the structural organization and behavior of such networks under different failure scenarios allows for the identification of critical elements and evaluation of potential targets for therapeutic intervention \cite{lasso2019structure,zhou2020network}.

The methodology employed in this study is rooted in the broader framework of statistical physics and complex network theory, which has proven effective in revealing hidden patterns of interaction in diverse systems. In particular, the ideas of network universality, structural robustness, and emergent behavior in collective dynamics were successfully applied in domains ranging from sociocultural phenomena to decision-making on complete graphs \cite{sarkanych2024consensus,sarkanych2016universality,holovatch2018statistical}. These foundational studies underscore the relevance of topological descriptors and network-based modelling as powerful tools for understanding both natural and artificial systems, including the virus--host interaction networks explored here.

In this work, we examine the main topological characteristics of viral networks, analyze their robustness to node removal, and assess the influence of the most central components on the global architecture of interactions.
The results obtained may contribute to a deeper understanding of the mechanisms underlying viral resilience and support the development of new approaches to combating viral infections.

\section{Data and methods}
\subsection{ViRBase database}

All data were obtained from the open access ViRBase v3.0 database \cite{virbase3.0:www},
which contains experimentally validated interactions between viral and host non-coding RNAs (ncRNAs).
ViRBase provides structured information on virus--host interactions, covering a broad range of species.
Table \ref{tab:virbase_statistics} presents statistics on the number of records in ViRBase for different host species.

\begin{table}[ht]
\caption{Distribution of records in ViRBase v3.0 by host species.}
\label{tab:virbase_statistics} 
\vspace{2pt}
    \centering
    \begin{tabular}{l|r}
        \hline\noalign{\smallskip}
        \textbf{Host species} & \textbf{Number of records}\\
        \noalign{\smallskip}\hline\noalign{\smallskip}
        Homo sapiens   & 710\,279 \\
        Mus musculus   &  67\,755 \\
        Sus scrofa     &  31\,595 \\
        Gallus gallus  &  17\,530 \\
        Zea mays       &   1545 \\
        \ldots\\
        \noalign{\smallskip}\hline\noalign{\smallskip}
        {\bf Total:} & 829\,338 \\
        \noalign{\smallskip}\hline
    \end{tabular}
\end{table}

ViRBase was selected for this study due to its relevance, extensive coverage, and open accessibility. It is widely used in virus--host interaction research and serves as a reliable data source for bioinformatics analyses \cite{li2015virbase,cheng2022virbase}.

This database contains structured information on interactions between viruses and host cell molecules, presented in tabular form.
Each record describes a specific interaction between a viral and a host component, including parameters such as virus species, its taxonomic classification, host species, the type of interacting molecule, and a corresponding reference to a scientific publication.
Table \ref{tab:virbase_sample} provides examples of such records.

\begin{table}[ht] 
\caption{Example of records from the ViRBase v3.0 database.}
\label{tab:virbase_sample}
\vspace{2pt}
    \scriptsize
    \centering
    \setlength{\tabcolsep}{3pt}
    \begin{tabular}{l l l l l l l l l} 
    \hline\noalign{\smallskip} 
        \textbf{ViRBase ID} & 
        \textbf{Virus Name} & 
        \textbf{Host Species} & 
        \textbf{Interactor1} & 
        \textbf{Interactor2} & 
        \textbf{PMID} & 
        \textbf{Score} \\ 
    \noalign{\smallskip} \hline\noalign{\smallskip} 
        HHID00000001 & Potato virus X & Nicotiana benthamiana & miRNA (nbe-miR166h-p5) & mRNA (KIP) & 30011130 & 0.9489 \\
        HHID00000004 & PRRSV & Homo sapiens & miRNA (hsa-miR-181b-5p) & mRNA (CD163) & 23740977 & 0.9526 \\
        HHID00000005 & PRRSV & Homo sapiens & miRNA (hsa-miR-125b-5p) & mRNA (NFKB2) & 23409058 & 0.8808 \\
        VVID00000248 & HIV-1 & Homo sapiens & miRNA (hiv1-miR-TAR-3p) & protein (tat) & 9485463 & 0.9820 \\
        VVID00000249 & Semliki Forest 4 & Homo sapiens & miRNA (miRNA sfv4-miRT124) & mRNA (SFVgp1) & 23077310 & 0.8808 \\ 
    \noalign{\smallskip} \hline 
    \end{tabular} 
\end{table}

The ``Score'' is a confidence metric prioritizing experimental evidence over computational predictions and penalizing ``weak’’ methods. It utilizes a weighted formula --- incorporating specific factors for strong experimental ($W_1 = 1.0$), weak experimental ($W_2 = 0.65$), and predicted ($W_3 = 0.25$) data --- to ensure that interactions supported by multiple independent resources receive significantly higher scores:
\begin{align}
\textrm{Score} = 1 - \prod_{i} \left(1 - \frac{W_i}{1+\re^{-x}}\right),
\end{align}
where $x$ is the number of resources \cite[see Q8 under Help]{virbase3.0:www}.

Note that interactions are defined at the molecular level (e.g., specific miRNA--mRNA pairs). Thus, multiple records for a single species pair represent distinct molecular pathways, not duplicated or redundant data.
Although the primary focus in ViRBase v3.0 is on virus--host interactions, the database also contains interactions between entities of the same type (virus--virus and host--host). These are included to preserve the full interaction context.

For the purposes of this study, three species were selected for further analysis: \textsl{Homo sapiens} (human), \textsl{Mus musculus} (house mouse), and \textsl{Gallus gallus} (chicken)\footnote{To be specific, \textsl{Gallus gallus} refers to red junglefowl, whereas chicken, as its domesticated subspecies, is \textsl{Gallus gallus domesticus}.}.
To ensure high confidence in the virus--host interaction data extracted from the database, we consider only the records with the highest available scores. 
Specifically, we select a threshold score just below the point where the number of records increases sharply.
This approach allows us to retain the most reliable interactions while avoiding the inclusion of potentially spurious or weakly supported data having lower confidence levels.
While the third most represented species is \textsl{Sus scrofa} (wild boar), the corresponding part of the database contains only five virus types --- considerably fewer than the two or three dozen found in mice and chickens. This means that a network built for \textsl{Sus scrofa} would have a substantially different structure from those of the other analyzed species, and is therefore not suitable for comparative analysis.

\subsection{Rank--frequency analysis of host biomolecule interactions with viruses}

To quantify the interaction activity of host biomolecules with viral agents, we calculated the number of interactions (bonds) for each biomolecule and ranked them in descending order of frequency. This allows us to identify the elements most involved in the network, potentially indicating their biological significance. Rank--frequency distributions were constructed for host biomolecule--virus interaction networks of \textsl{Homo sapiens} under different interaction confidence thresholds. Empirical frequencies were compared with theoretical models, including Zipf's law, truncated Gaussian, and Yule distributions, in order to assess the presence of scaling behavior and truncation effects. This approach allows us to evaluate the robustness of the observed rank--frequency patterns with respect to interaction confidence filtering and dataset size.

Table~\ref{tab:rank_freq_table} presents the top-ranking host biomolecules based on two filtering thresholds for interaction confidence scores: $\text{Score} \geqslant 0.95$ and $\text{Score} > 0$. For each threshold, rank $r$, symbol, and frequency $f_r$ are listed.

\begin{table}[ht]
	\caption{Top host biomolecules by number of virus interactions for two threshold values.}
	\label{tab:rank_freq_table}
	\vspace{2pt}
    \centering
    \begin{tabular}{c l c | l c}
        \hline\noalign{\smallskip} 
        \multirow{2}{*}{$\bm r$} & \multicolumn{2}{c|}{$\textbf{Score} \bm{\geqslant 95}$} & \multicolumn{2}{c}{\bf All scores} \\
        \cmidrule(lr){2-3}
        \cmidrule(lr){4-5}
        & {\bf Symbol} & $\bm{f_r}$ & {\bf Symbol} & $\bm{f_r}$ \\
        \noalign{\smallskip}\hline\noalign{\smallskip} 
        1 & hsa-miR-155-5p & 62 & hsa-miR-27b-5p & 3398 \\ 
        2 & hsa-miR-146a-5p & 32 & hsa-miR-32-5p & 1684 \\ 
        3 & hsa-miR-19a-3p & 26 & hsa-miR-190a-5p & 1645 \\ 
        4 & hsa-miR-17-5p & 24 & hsa-miR-190a-3p & 1592 \\ 
        5 & hsa-miR-93-5p & 20 & hsa-miR-4521 & 1576 \\ 
        6 & hsa-miR-181b-5p & 18 & hsa-miR-365a-5p & 1541 \\ 
        7 & hsa-miR-92a-3p & 18 & hsa-miR-142-3p & 1506 \\ 
        8 & hsa-miR-203a-3p & 18 & hsa-miR-155-5p & 1378 \\ 
        9 & DICER1 & 18 & hsa-miR-17-5p & 1298 \\ 
        10 & MYC & 18 & hsa-miR-20a-5p & 1288 \\ 
        11 & 7SL & 16 & hsa-miR-93-5p & 1286 \\ 
        12 & hsa-miR-149-5p & 16 & hsa-miR-20b-5p & 1280 \\ 
        \multicolumn{5}{c}{\quad\ \dots}\\
        \noalign{\smallskip}\hline
    \end{tabular}
\end{table}

The corresponding rank--frequency dependencies exhibit a Zipf-like behavior. Zipf's law is a classical statistical regularity observed in many complex systems, where the frequency $f(r)$ of an element is inversely proportional to its rank $r$ in a sorted list \cite{zipf2013psycho}:

\begin{equation}
    f(r) = \frac{C}{r^s},
\end{equation}
with $C$ and $s$ being positive constants. Originally formulated for linguistics, this law has been successfully applied to a wide range of systems, including biological networks, highlighting the self-organized nature and hierarchy among elements such as genes, proteins, or regulatory RNAs \cite{kalankesh2012language,rovenchak2018telling,semple2022linguistic}. However, deviations from a pure Zipf-like behavior are commonly observed in finite and biologically constrained systems, motivating the consideration of truncated distributions.

In our case, the distribution of interaction frequencies among host biomolecules demonstrates a Zipf-like behavior, particularly at lower ranks (i.e., most active biomolecules). However, the tail of the distribution exhibits a sharper decay than predicted by the pure power law. 
To better capture this effect, we employ a power-law model with a stretched-exponential (Gaussian-type) cutoff, commonly used to account for finite-size effects in empirical rank--frequency distributions, cf.~\cite{laherrere1998stretched}:
\begin{equation}
    f(r) = \frac{C}{r^s} \cdot \re^{-\left(r / r_0\right)^2},
\end{equation}
where the exponential factor introduces a soft cutoff beyond rank $r_0$, reflecting saturation or sampling limits in real biological systems.
In what follows, we refer to this model as the Gaussian-truncated Zipf's law.

In addition to the above models, we also considered a Yule-type (geometrically truncated power-law) rank model:
\begin{equation}
    f(r) = A\frac{k^r}{r^b}.
\end{equation}
The Yule form is well established in quantitative linguistics as a flexible description of rank--frequency statistics of texts, cf.~\cite{rovenchak2018diary}.

\begin{figure}[ht]
    \centering
    \includegraphics[scale=0.5]{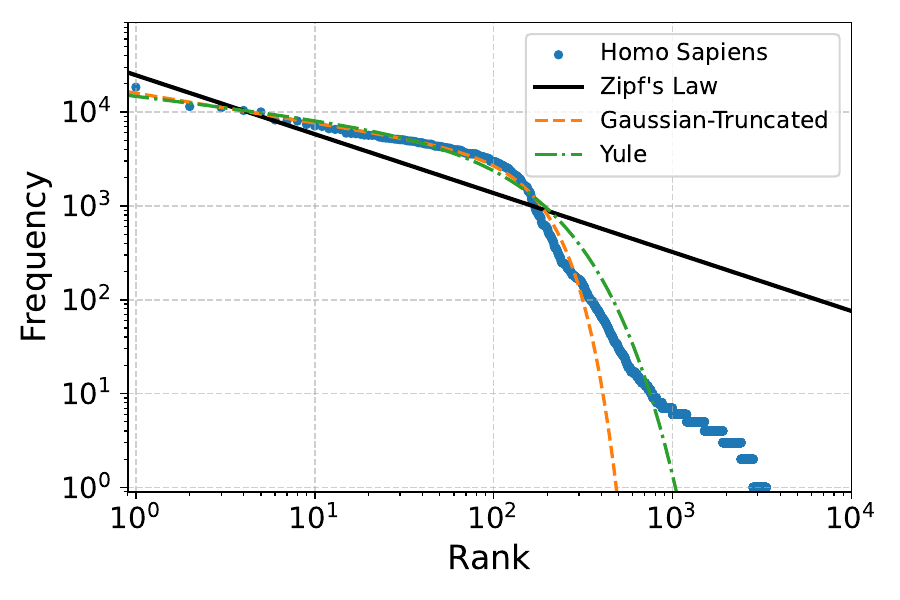}
    \caption{(Colour online) Rank--frequency distribution for all interactions (all scores).}
    \label{fig:zipf_all}
\end{figure}

\begin{figure}[ht]
	\centering
	\includegraphics[scale=0.5]{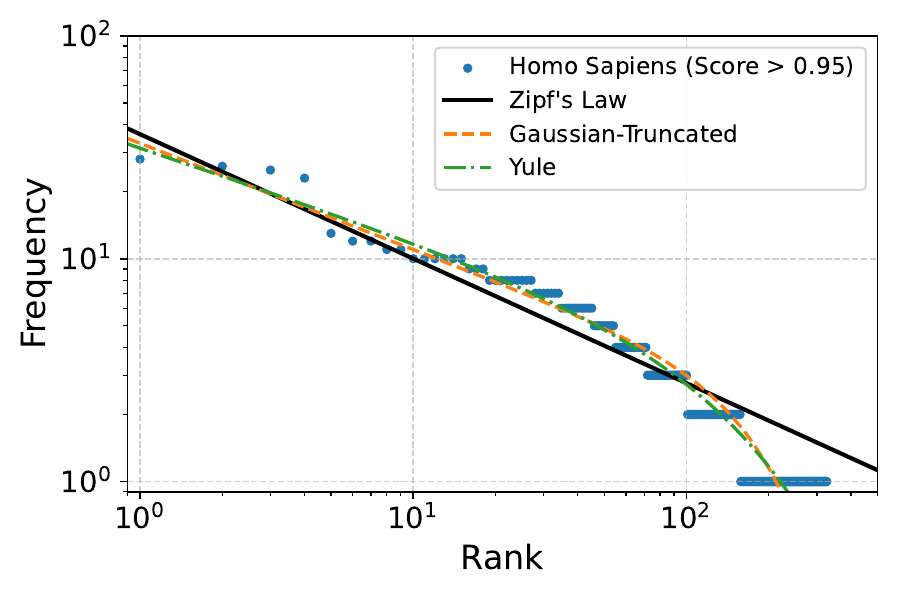}
	\caption{(Colour online) Rank--frequency distribution for high-confidence human biomolecule interactions ($\text{Score} \geqslant 0.95$).}
	\label{fig:zipf_95}
\end{figure}

\begin{figure}[ht!]
	\centering
	\includegraphics[scale=0.5]{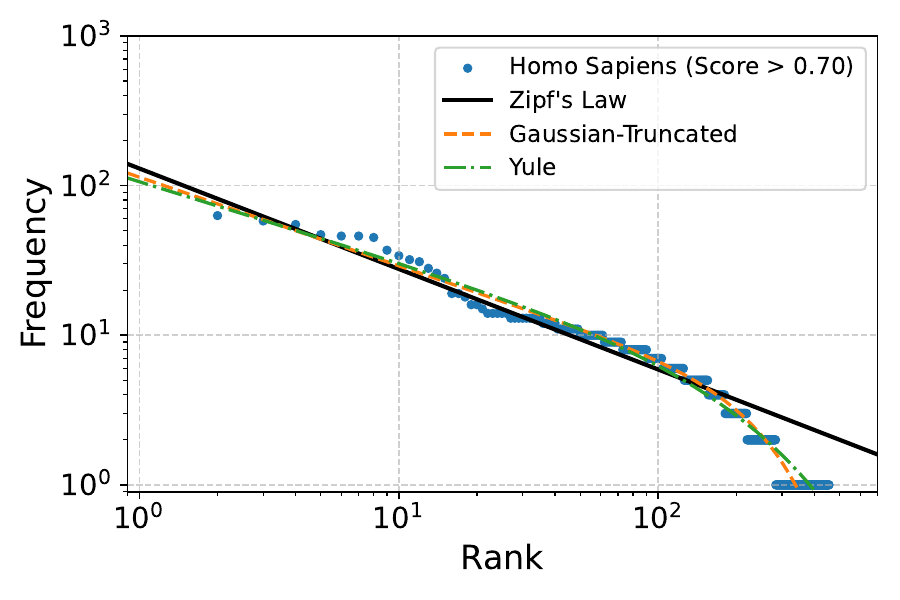}
	\caption{(Colour online) Rank--frequency distribution for high-confidence human biomolecule interactions ($\text{Score} \geqslant 0.70$).}
	\label{fig:zipf_70}
\end{figure}

Figures~\ref{fig:zipf_all},~\ref{fig:zipf_95}, and~\ref{fig:zipf_70} illustrate the rank--frequency distributions for different interaction confidence thresholds. While lower thresholds yield broader distributions with heavier tails, higher confidence filtering results in a clearer truncation at large ranks.
The fitting parameters for both Zipf and Gaussian-truncated Zipf models are presented in table~\ref{tab:fit_params}.

We restricted the analysis in this section to host--biomolecule interactions in \textit{Homo sapiens} because the datasets available for the other two species were too small to support stable parameter estimation and robust goodness-of-fit assessments. As expected, stricter filtering by interaction score improved the fit quality across models, with the highest-threshold subset (Score $>0.95$) showing the clearest agreement, particularly for the Yule and Gaussian-truncated laws. Such truncation effects may reflect biological constraints, including the limited number of host biomolecules capable of sustaining a large number of viral interactions.

The rank--frequency distribution shown in figure~\ref{fig:zipf_all} exhibits a multi-regime structure, with several approximately linear segments in log--log coordinates. Although this behavior is not directly equivalent to the rank--frequency relations in linguistic systems, a formal analogy may be drawn with the studies where deviations from a single Zipf's law were associated with the coexistence of functionally distinct classes of elements  \cite{ferrer-i-cancho_sole2001,buk_rovenchak2004,piantadosi2014,holovatch_palchykov2016,rovenchak_buk2018}.

\begin{table}[ht]
	\caption{Fitting parameters for rank--frequency models.}
	\label{tab:fit_params}
	\vspace{2pt}
	\centering
	\setlength{\tabcolsep}{6pt}
	\renewcommand{\arraystretch}{1.15}
	\begin{tabular}{l cc ccc ccc}
		\toprule
		\multirow{2}{*}{\bf Dataset} & 
		\multicolumn{2}{c}{\bf Zipf} & 
		\multicolumn{3}{c}{\bf Gaussian-truncated} & 
		\multicolumn{3}{c}{\bf Yule} \\
		\cmidrule(lr){2-3}\cmidrule(lr){4-6}\cmidrule(lr){7-9}
		& $\bm C$ & $\bm s$ & $\bm C$ & $\bm{s}$ & $\bm{r_0}$ & $\bm A$ & $\bm k$ & $\bm b$ \\
		\midrule
		Score $> 0.95$ & 36.22 & 0.5586 & 32.97 & 0.4758 & 218.7 & 31.52 & 0.9944 & 0.4090 \\
		Score $> 0.70$ & 130.0 & 0.6716 & 113.8 & 0.5914 & 297.5 & 106.5 & 0.9961 & 0.5318 \\
		All (no threshold) & 24699 & 0.6281 & 15912 & 0.3167 & 174.6 & 14822 & 0.9923 & 0.2316 \\
		\bottomrule
	\end{tabular}
\end{table}

This analogy, however, should be treated with caution. In contrast to linguistic corpora, where rank reflects the token frequency and datasets are typically very large, the present distributions are derived from structural network properties (node degree) and are based on comparatively smaller samples. As a result, the statistical resolution is limited, and the mechanisms underlying Zipf-like behavior cannot be directly inferred.

Instead, the presence of multiple (in this case, three) scaling regimes indicates a heterogeneous organization of the network, probably reflecting different classes of nodes such as hubs, intermediate nodes, and peripheral elements. A similar interpretation is already less clear for datasets with specified confidence scores, as shown in figures~\ref{fig:zipf_95} and \ref{fig:zipf_70}.
In these cases, the rank–frequency distributions resemble the syllable-level rather than the word-level behavior, showing a steeper decline at high ranks, cf.~\cite{rovenchak2018diary}.

\subsection{Building a network}

\begin{figure}[ht]
    \centering
    \includegraphics[width=0.48\textwidth]{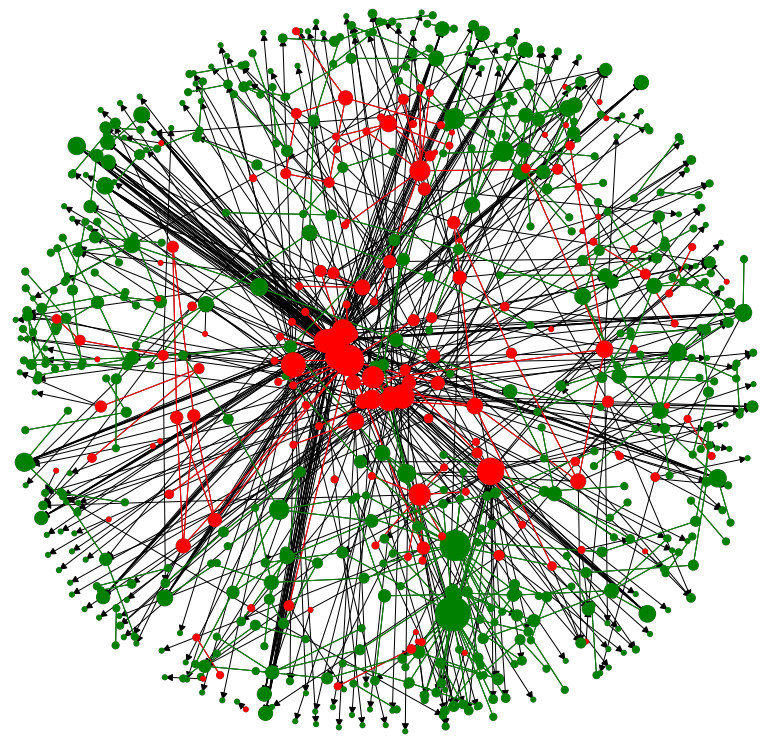}
    \caption{(Colour online) Virus--host network for {\sl Homo sapiens}.}
    \label{fig:network_hs}
\end{figure}

We represent the nodes of the network as biomolecules, primarily microRNAs or proteins. For clarity, nodes corresponding to viral biomolecules are shown in red, while those corresponding to host biomolecules are shown in green. Virus--host connections are illustrated in black, assigned a weight of $w_1 = 1$, and are considered to be directed. Indirect interactions --- virus--virus (red) and host--host (green) --- are incorporated as undirected edges and assigned lower weights $\left(w < 1\right)$, reflecting their secondary role relative to cross-species interactions. Technically, we set $w=0.5$ to obtain all the data reported below.
Figure~\ref{fig:network_hs} shows an example of such a network for \textsl{Homo sapiens}; its characteristics, together with the data for two other species, are given in table~\ref{tab:network_params}. 
The network is not fully strongly connected; therefore, global path-based metrics are evaluated on the largest strongly connected component (SCC). Additionally, properties of the largest weakly connected component (WCC) are reported to ensure robustness.

An analysis of the node strength distribution (i.e., the sum of weights over both incoming and outgoing edges per node) reveals that its rank--strength representation follows a power-law dependence (Zipf-like law):
\begin{align} \label{eq:zipf}
    s(r) \propto r^{-\zeta},
\end{align} 
where $r$ denotes the rank and $\zeta$ is the corresponding exponent.
We fitted this law to three sets of nodes, namely, virus-associated nodes, host-associated nodes, and all nodes combined, disregarding the first $20\%$ of data points.
The resulting exponents $\zeta_{\rm virus}$, $\zeta_{\rm host}$, and $\zeta_{\rm all}$ are presented in table~\ref{tab:network_params}.
Note that this representation differs from the probability distribution $P(k)$ of node strength (or degree), which is commonly used in network analysis.

\begin{table}[ht]
	\centering
	\caption{Network properties for virus--host systems.}
	\label{tab:network_params}
	\begin{tabular}{lccc}
		\hline\noalign{\smallskip}
		& \textbf{\textit{Homo sapiens}} & 
		\textbf{\textit{Gallus gallus}} & 
		\textbf{\textit{Mus musculus}} \\
		\noalign{\smallskip}\hline\noalign{\smallskip}
		Total nodes                  & 715   & 191   & 115   \\
		Virus nodes                  & 157   & 35    & 26    \\
		Host nodes                   & 558   & 156   & 89    \\
		Total edges                  & 868   & 209   & 87    \\
		Directed edges               & 485   & 127   & 29    \\
		Undirected edges             & 383   & 82    & 58    \\
		Density                      & 0.00245 & 0.00802 & 0.0111 \\
		Diameter                     & 20    & 11    & 4     \\
		Average Path Length          & 6.42  & 4.95  & 2.39  \\
		Mean Distance in largest SCC & 5.03  & 2.85  & 2.39  \\
		Mean Distance in largest WCC & 6.42  & 4.95  & 2.39  \\
		Assortativity                & 0.070 & $-0.245$ & $-0.389$ \\
		Betweenness                  & 0.0134 & 0.00844 & 0.00403 \\
		$\zeta_{\text{all}}$        & 0.926 & 0.853 & 0.569 \\
		$\zeta_{\text{virus}}$      & 1.39  & 1.74  & 0.258 \\
		$\zeta_{\text{host}}$       & 0.747 & 0.601 & 0.609 \\
		\noalign{\smallskip}\hline
	\end{tabular}
\end{table}

For an ideal power-law distribution $P(k) \propto k^{-\gamma}$, the corresponding rank--strength exponent is related as $\zeta = 1/(\gamma - 1)$, cf.~\cite{simon1955,newman2005}. However, given the limited size of the dataset and the presence of multiple scaling regimes, this relation should be interpreted with caution.

\begin{figure}[h]
\centering
\begin{tabular}{cc}
{\includegraphics[clip,scale=0.136]{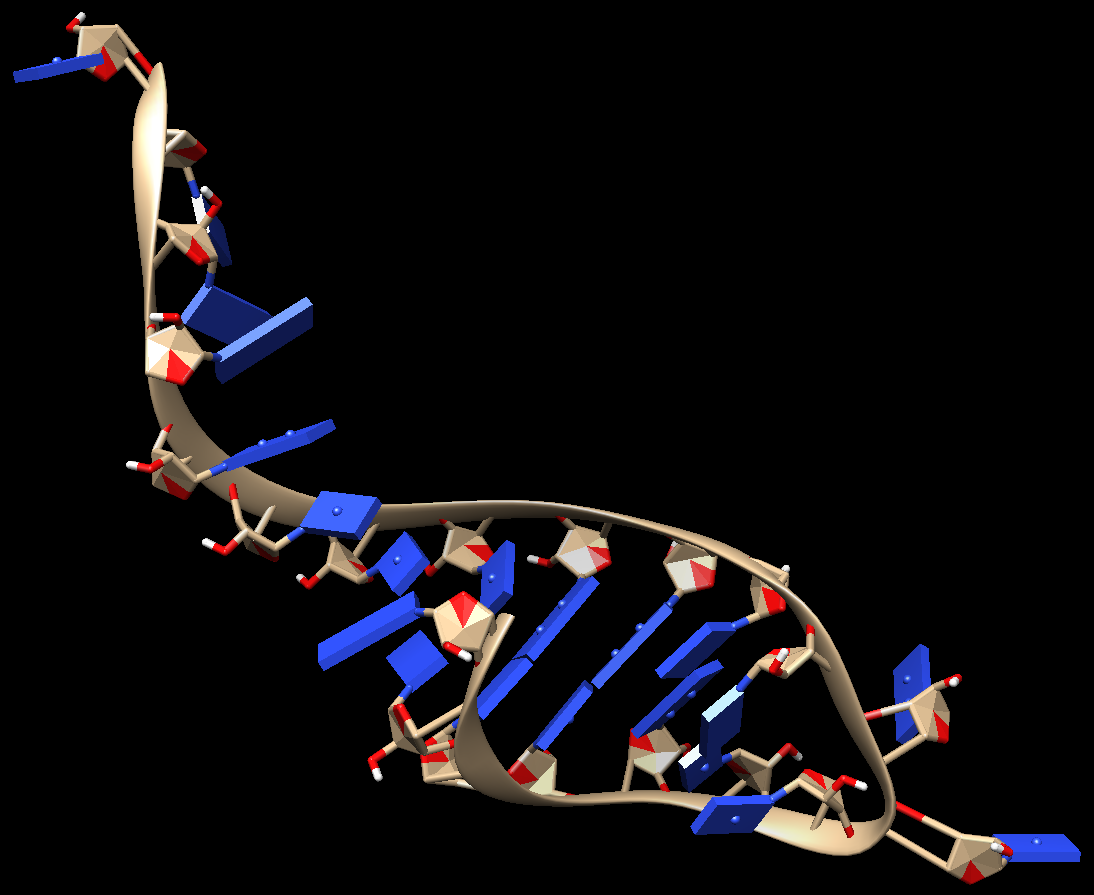}} & 
{\includegraphics[clip,scale=0.136]{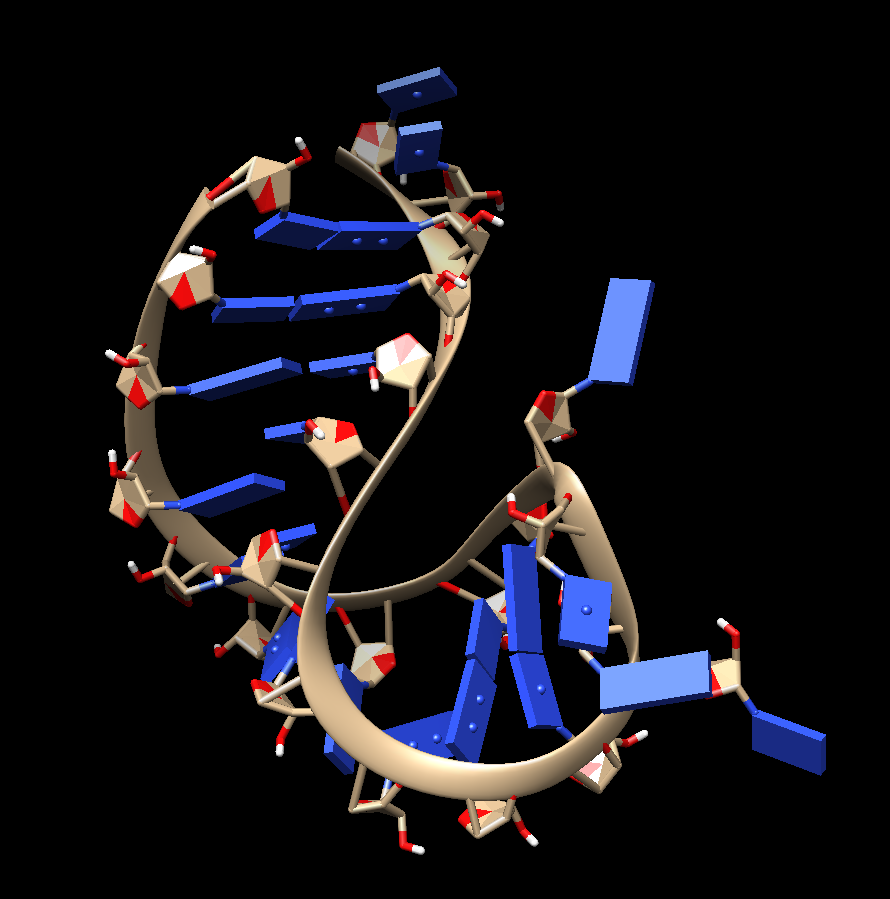}} \\
(a) & (b)
\end{tabular}
\caption{(Colour online) Biomolecules corresponding to the most influential nodes in the \textsl{Homo sapiens} network:
(a)~MicroRNA hcmv-miR-US25-1-5p of the Human Cytomegalovirus,
(b)~Human microRNA hsa-miR-155-5p (proinflammatory, oncogenic).
Molecular images were generated using UCSF Chimera 1.18, developed by the Resource for Biocomputing, Visualization, and Informatics at the University of California, San Francisco (supported by NIH P41-GM103311) \cite{Pettersen_etal:2004}.}
\label{fig:top-images}
\end{figure}

Detailed information about the top strongest nodes is shown in table~\ref{tab:top_nodes_hs} for the \textsl{Homo sapiens} network.
The most influential nodes are shown in figure~\ref{fig:top-images}.
The host biomolecules collectively participate in innate and adaptive immune responses, inflammatory signaling, and oncogenic processes, reflecting the complex interplay between antiviral defense and the pathways that viruses exploit for replication and persistence.

\begin{table}[h!]
	\begin{center}
		\caption{Top nodes by strength at 75\% and 95\% confidence score thresholds.}
		\label{tab:top_nodes_hs}
		{\footnotesize
			\begin{tabular}{l|r|l|r|l}
				\hline\noalign{\smallskip}
				\textbf{Node (75\%)} & \textbf{Strength} & \textbf{Node (95\%)} & \textbf{Strength} & \textbf{Name (for 95\%)} \\
				\noalign{\smallskip}\hline\noalign{\smallskip}
				\multicolumn{5}{c}{\textbf{Virus}} \\
				\noalign{\smallskip}\hline\noalign{\smallskip}
				hcmv-miR-US25-1-5p & 682.0 & kshv-miR-K12-11-3p & 28.0 & Kaposi sarcoma-associated herpesvirus \\
				ebv-miR-BART18-5p  & 46.0  & PB1                & 23.0 & Influenza polymerase basic 1 \\
				X                  & 41.0  & ebv-miR-BART18-5p  & 23.0 & Epstein Barr virus \\
				kshv-miR-K12-11-3p & 37.0  & hcmv-miR-US25-1-5p & 22.0 & Human cytomegalovirus \\
				NS1                & 24.0  & NS1                & 20.0 & Influenza nonstructural protein 1 \\
				NP                 & 24.0  & M1                 & 20.0 & Influenza matrix protein 1 \\
				PB1                & 24.0  & NP                 & 20.0 & Influenza nucleoprotein \\
				M1                 & 23.0  & PA                 & 19.0 & Influenza polymerase acidic \\
				PA                 & 19.0  & PB2                & 19.0 & Influenza polymerase basic 2 \\
				PB2                & 19.0  & X                  & 18.0 & X protein [Hepatitis B virus] \\
				\noalign{\smallskip}\hline\noalign{\smallskip}
				\multicolumn{5}{c}{\textbf{Host}} \\
				\noalign{\smallskip}\hline\noalign{\smallskip}
				hsa-miR-30a-5p     & 28.5  & hsa-miR-155-5p     & 15.5 & (See explanation under the Table)\\
				hsa-miR-130a-3p    & 28.5  & hsa-miR-17-5p      & 10.5 & \\
				IFNB1 & & & &\\
				\quad(Interferon beta 1) & 27.0  & hsa-miR-19a-3p     & 10.0 & \\
				hsa-miR-30d-5p     & 27.0  & hsa-miR-222-3p     & 9.5  & \\
				hsa-miR-30b-5p     & 23.5  & hsa-miR-16-5p      & 9.5  & \\
				hsa-miR-30c-5p     & 23.0  & hsa-miR-34a-5p     & 9.5  & \\
				hsa-miR-30d-3p     & 22.0  & hsa-miR-142-5p     & 9.0  & \\
				CASP3 (Caspase 3) & 18.5  & MYC                & 9.0  & MYC proto-oncogene\\
				& & & & bHLH transcription factor \\
				hsa-miR-155-5p     & 17.0  & hsa-miR-93-5p      & 9.0  & \\
				hsa-miR-766-3p     & 17.0  & hsa-miR-181b-5p    & 8.5  & \\
				\noalign{\smallskip}\hline
			\end{tabular}
		}
	\end{center}
	{\footnotesize 
		Explanation of microRNA names: \texttt{hsa} stands for \textsl{Homo sapiens}, \texttt{miR} indicates a microRNA (mature form, as opposed to precursor), the number is an identifier in the miRBase \cite{mirbase:www}, and \texttt{-5p} or \texttt{-3p} indicates that this miRNA comes from the $5'$ or the $3'$ arm of the precursor hairpin structure. 
		
	}
\end{table}

\begin{figure}[h]
	\centering
	
	\includegraphics[scale=0.5]{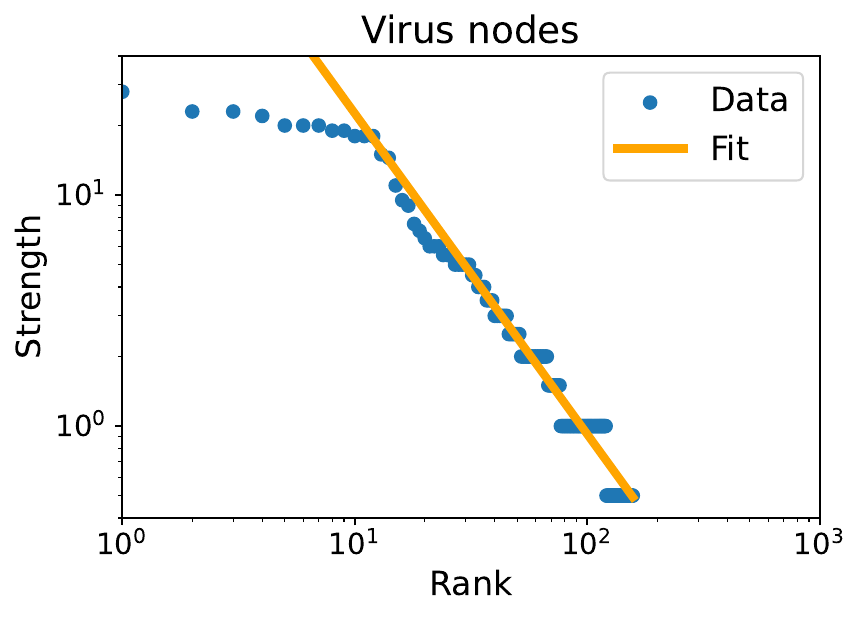}
	\includegraphics[scale=0.50]{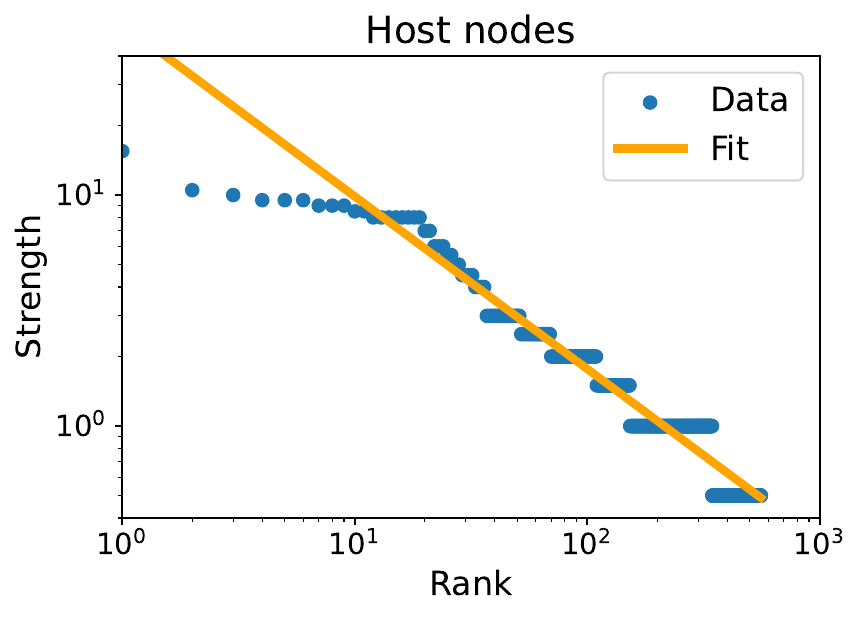}
	
	\caption{(Colour online) Node strength distribution for the virus--host network of \textsl{Homo sapiens}. The plot demonstrates a power-law behavior (\ref{eq:zipf}). The distribution is shown separately for virus-associated nodes and host-associated nodes. The power-law exponent $\gamma$ is estimated after discarding the lowest 20\% of values.}
\end{figure}


\section{Results}
\subsection{Network dynamics}

To investigate the  dynamic behavior of the network, we employed two strategies: sequential removal of the most influential nodes and random node deletion. By monitoring the resulting changes in the system, we obtain parameters that characterize its dynamic response.

Since the connections form a directed graph, the network density is defined in the standard way:
\begin{equation}
D = \frac{m}{n(n - 1)},
\end{equation}
\noindent where $m$ is the number of edges in the graph, $n$ is the number of nodes, and $n \left(n - 1\right)$  represents the maximum possible number of directed edges. In practice, this was computed using the {\tt networkx.density()} method.

The assortativity coefficient measures the tendency of the nodes with similar degrees (number of connections) to connect with each other. Its value ranges from $-1$ to $1$: positive values indicate that high-degree nodes are more likely to connect with other high-degree nodes, whereas negative values suggest that high-degree nodes tend to connect with low-degree nodes.

To study the structural properties of the virus--host interaction network, we define the adjacency matrix $A$ for the integrated system. For a network of $N$ nodes (comprising both viral and host molecules), the elements $A_{ij}$ represent the weight of the directed edge from biomolecule $i$ to biomolecule $j$:
\begin{align}
    A_{ij} = 
    \begin{cases} 
      w_1 & \text{if } i \to j \text{ is a direct virus--host interaction}, \\
      w/2 & \text{if } i \to j \text{ is an indirect intra-type interaction}, \\
      0 & \text{otherwise}.
    \end{cases}
\end{align}
As discussed previously, virus--host connections are assigned a weight of $w_1 = 1$, while indirect interactions --- specifically virus--virus (red) and host--host (green) connections --- are assigned a weight of $w$. Since these indirect interactions are considered bidirectional, both $A_{ij}$ and $A_{ji}$ are set to $w/2$ for such pairs, resulting in a total edge weight of $w$ between the connected biomolecules of the same type.
This adjacency matrix was computed in Python using the {\tt NetworkX} library; it serves as the basis for subsequent calculations, including temperature and magnetization.

In network analysis, one of the key parameters is the graph temperature, which can be defined via the eigenvalues of its adjacency matrix $A$ as:
\begin{equation}
    T = \frac{\langle (\Delta\lambda)^2\rangle}{\langle \lambda\rangle},
\end{equation}
where $\lambda_j$ are the eigenvalues of the adjacency matrix, $\langle \lambda\rangle$ is their mean, and $\langle (\Delta\lambda)^2\rangle$ is the mean squared fluctuation of the eigenvalues.

This expression resembles the definition of temperature in the theory of thermodynamic fluctuations, where temperature is proportional to the ratio of the mean square fluctuation of energy to its average value. In physical systems, this reflects the level of randomness and disorder, whereas in networks, a high temperature may indicate a high variability in connection structure, and a low temperature may correspond to a more ordered, structured organization \cite{estrada2012structure,bianconi2009entropy}.

To investigate the network dynamics, a step-by-step node removal procedure was applied, followed by an analysis of changes in its structural characteristics. At each iteration, a single node was removed, after which parameters such as the size of the largest connected component, average shortest path length, degree assortativity coefficient, and graph temperature were recalculated. This approach enables an assessment of the robustness of a network to structural perturbations, the identification of critical nodes maintaining its connectivity \cite{callaway2000network}, and the observation of patterns in the degradation of network topology~\cite{albert2000error}.

\begin{figure*}[h!]
	\centering
	\setlength{\tabcolsep}{2.6pt}
	\renewcommand{\arraystretch}{1.0}
	
	\begin{tabular}{ccc}
		\includegraphics[width=0.32\linewidth]{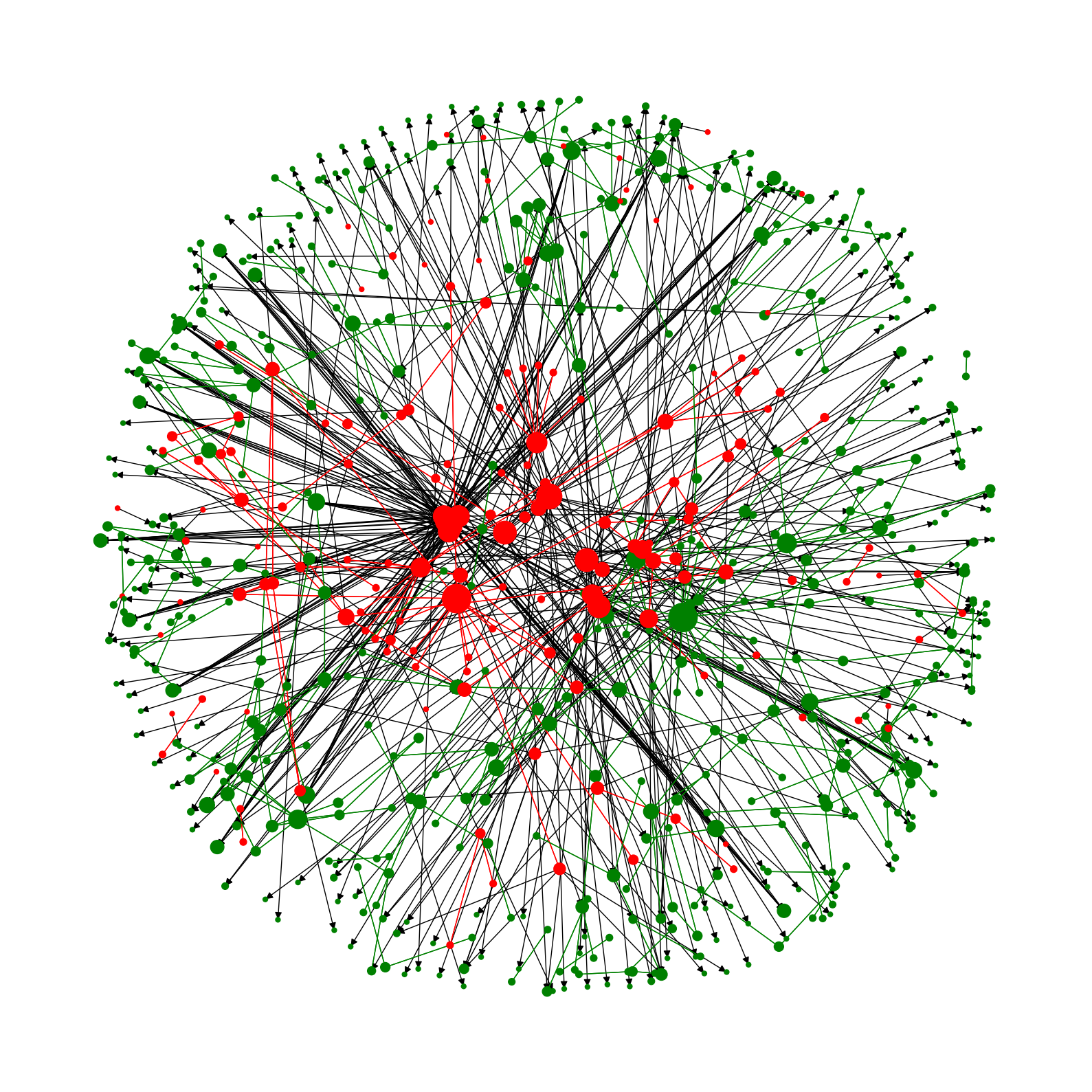} &
		\includegraphics[width=0.32\linewidth]{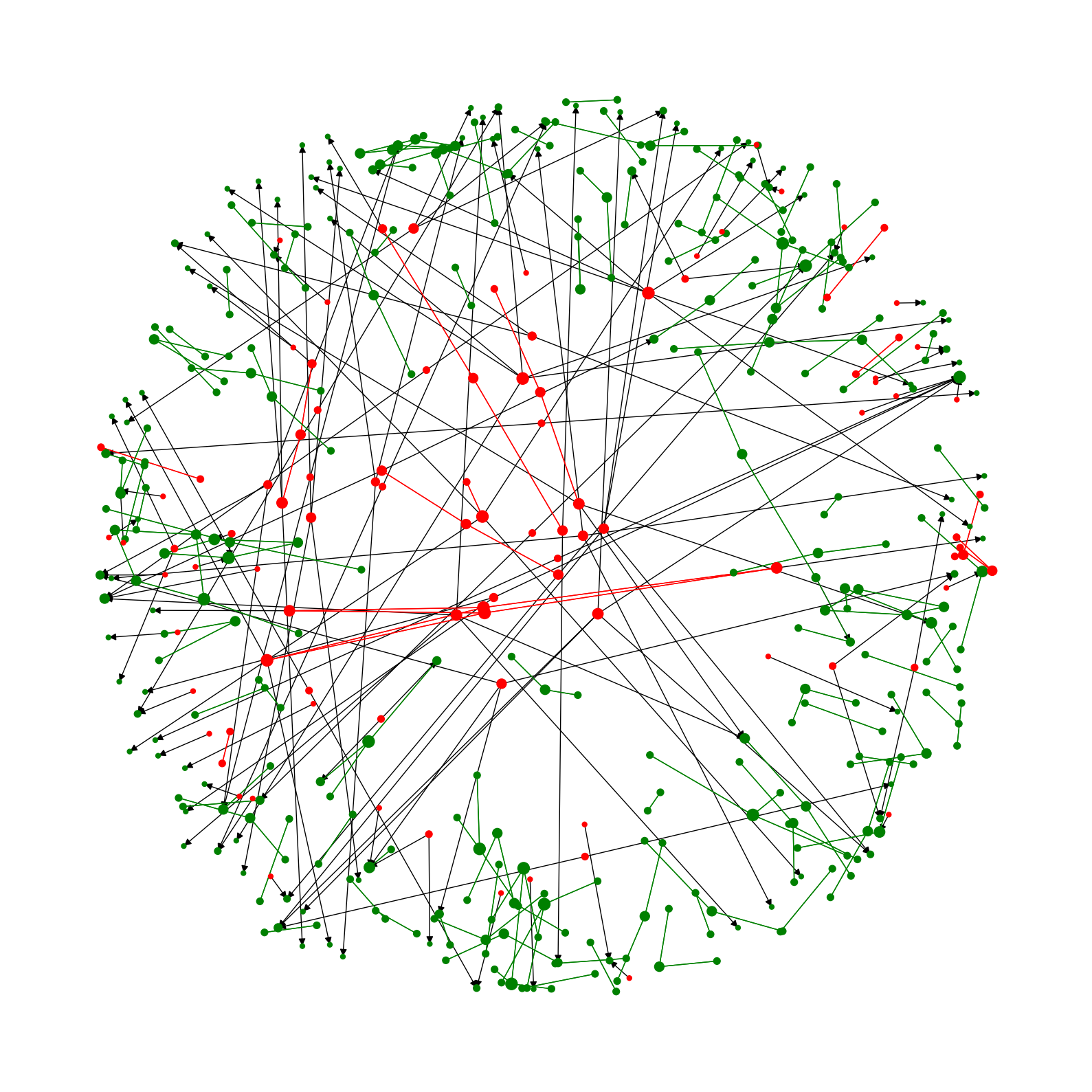} &
		\includegraphics[width=0.32\linewidth]{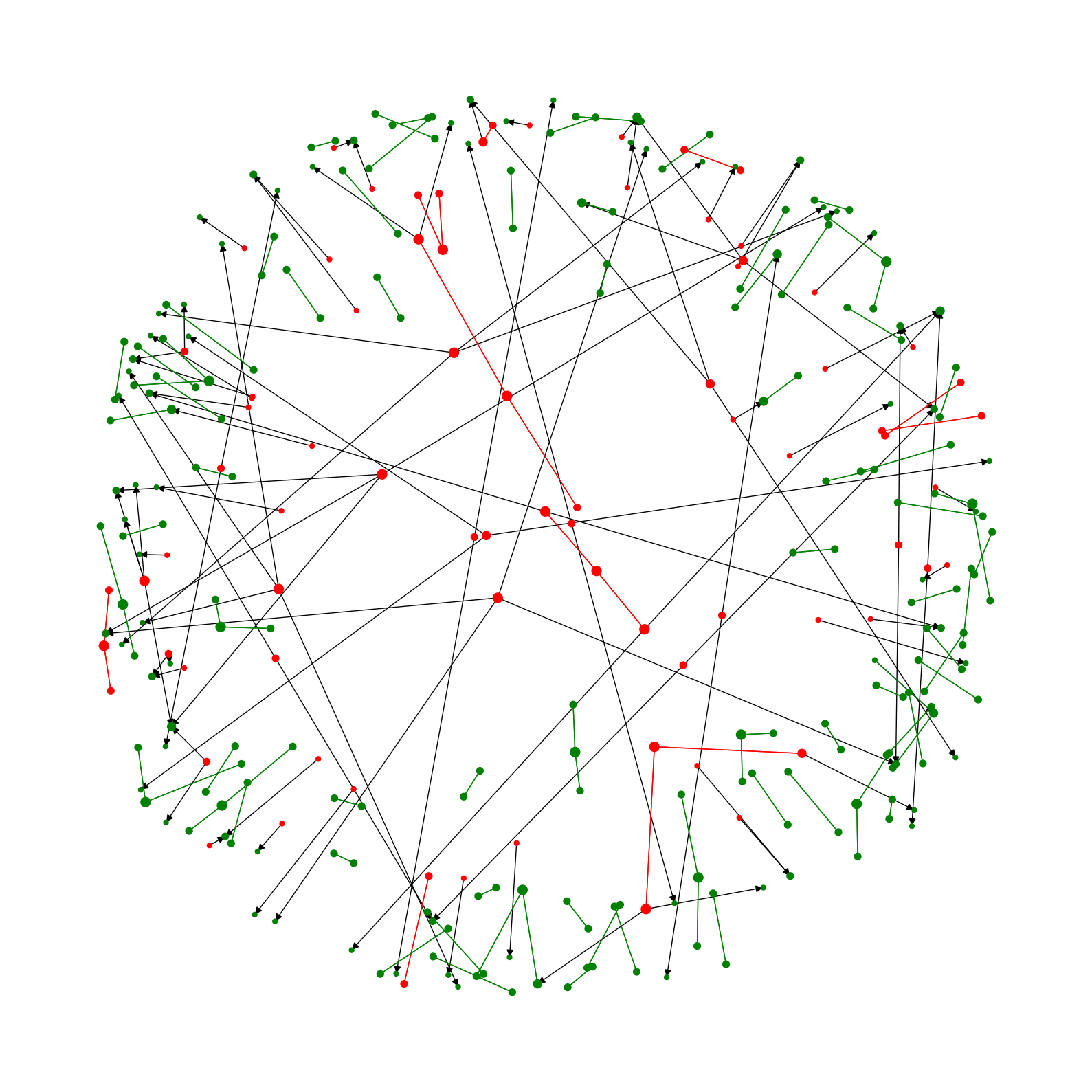} \\
		{\footnotesize step 000} & {\footnotesize step 050} & {\footnotesize step 100} \\
		\includegraphics[width=0.32\linewidth]{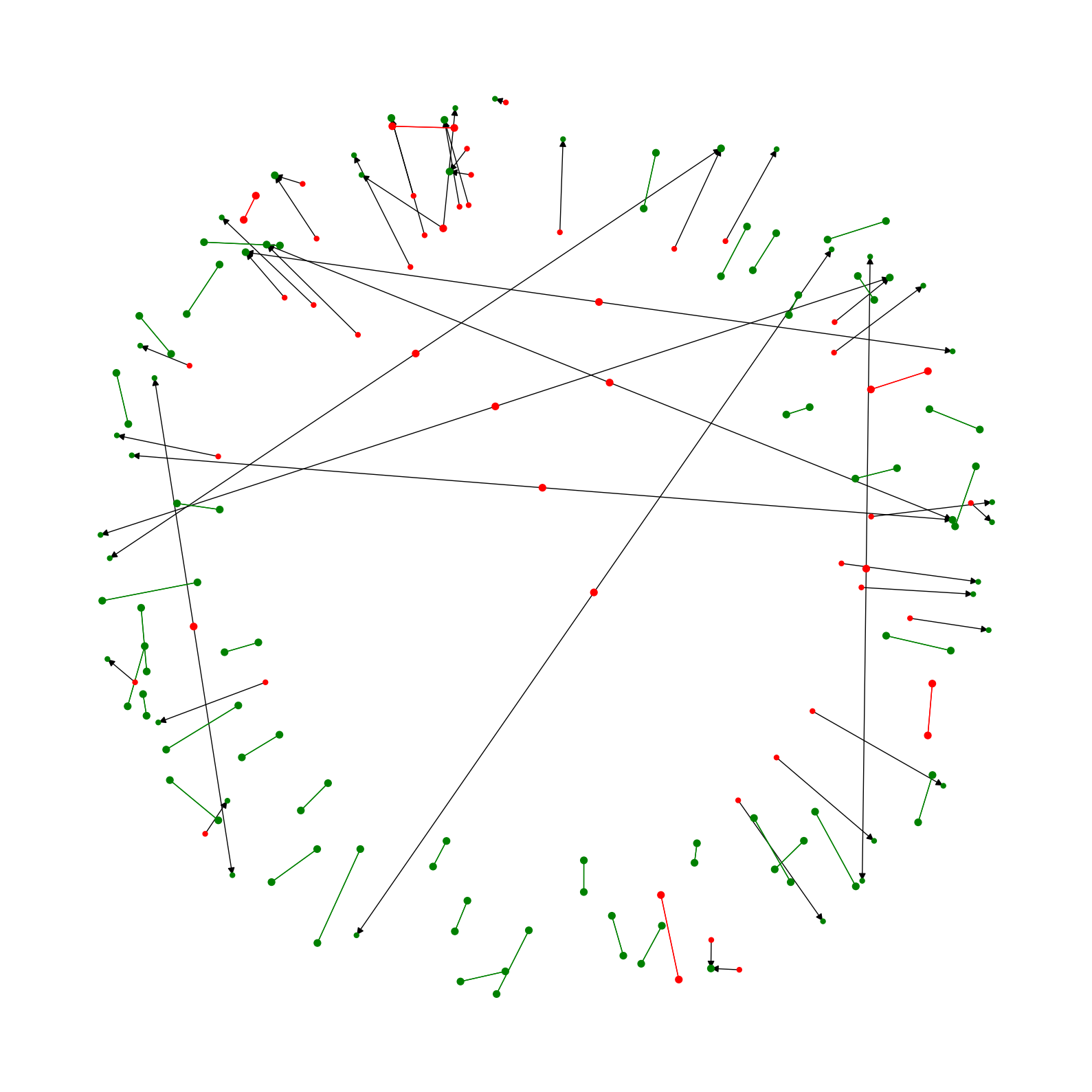} &
		\includegraphics[width=0.32\linewidth]{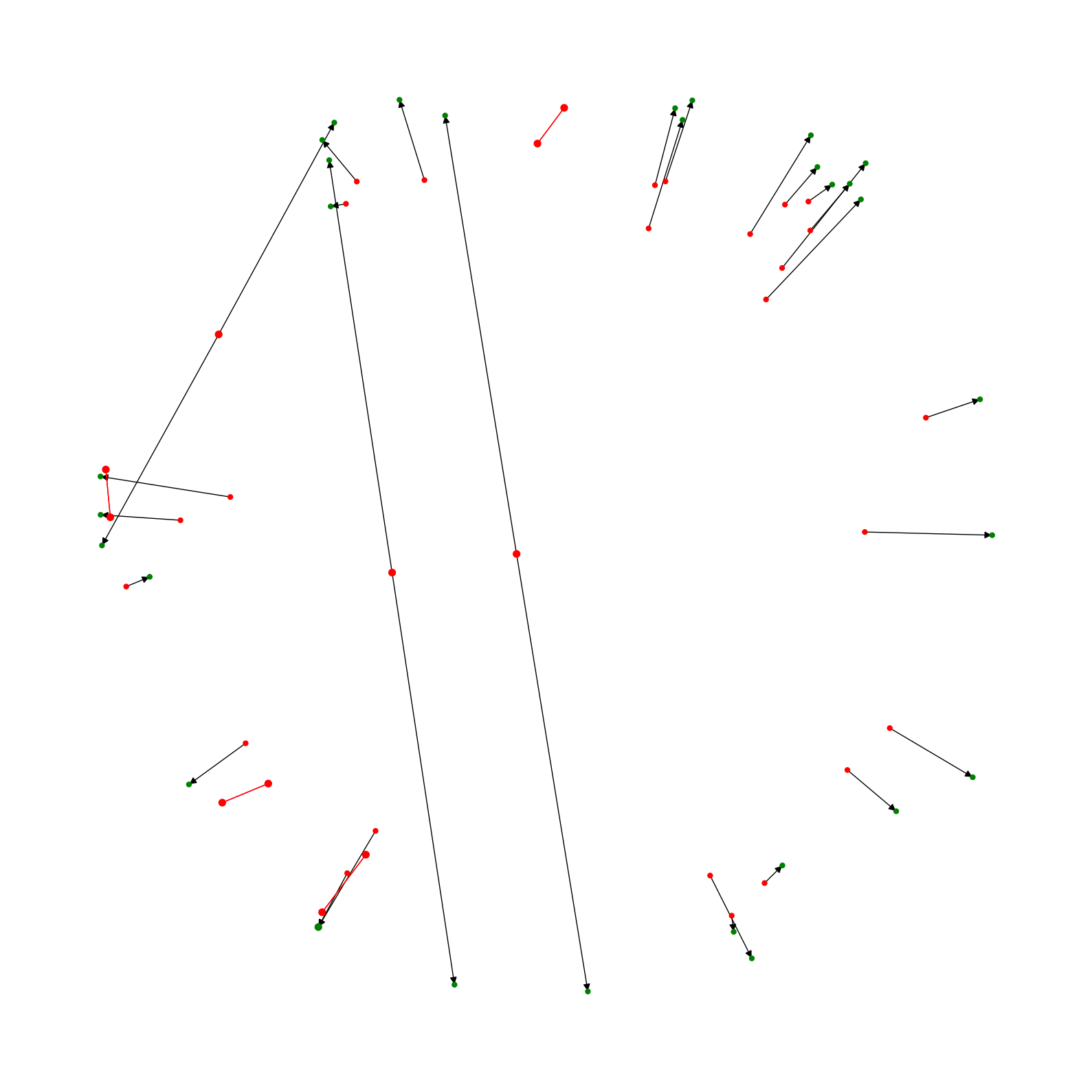} &
		\includegraphics[width=0.32\linewidth]{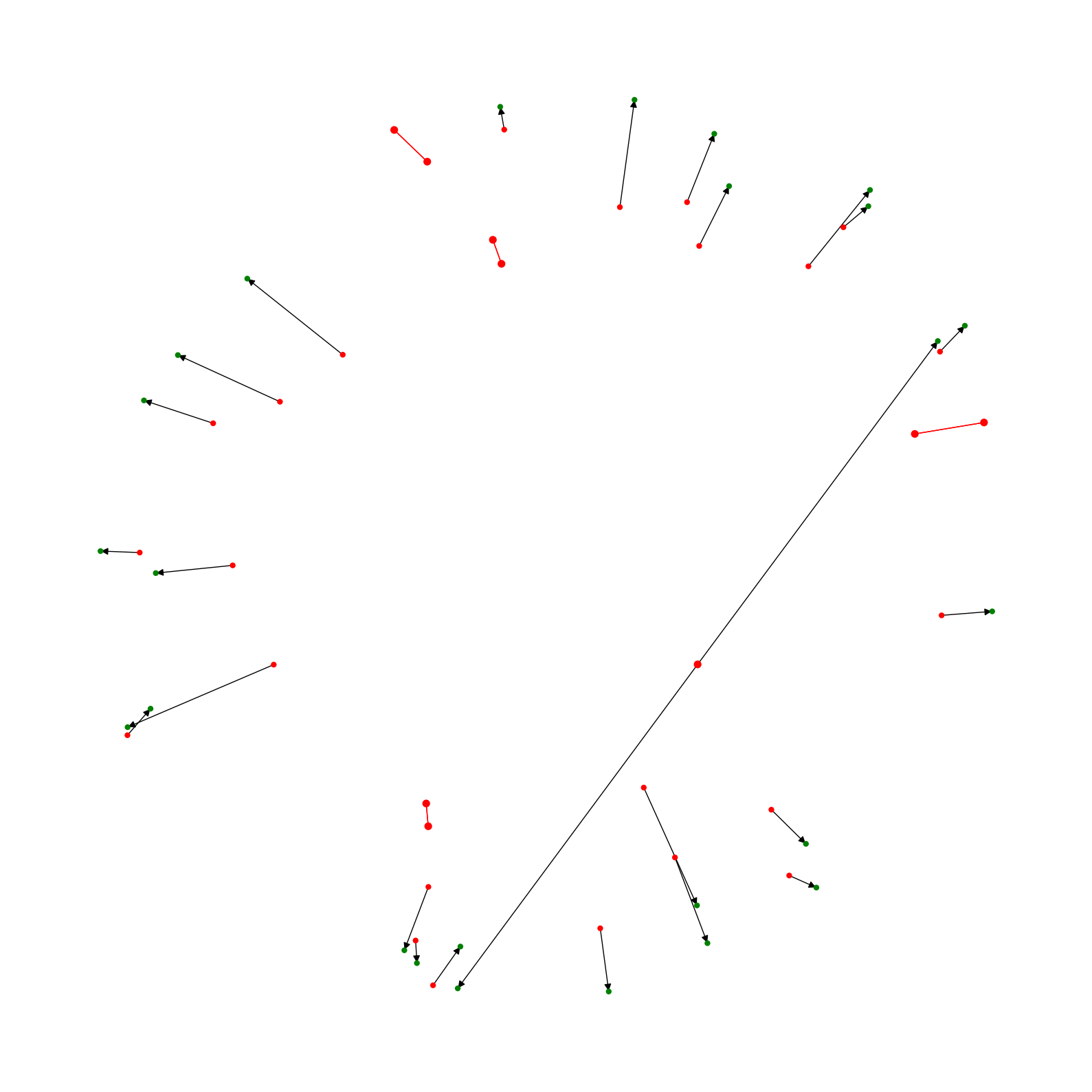} \\
		{\footnotesize step 150} & {\footnotesize step 200} & {\footnotesize step last} \\
	\end{tabular}
	
	\caption{(Colour online) Network evolution snapshots for \textsl{Homo sapiens} under targeted node removal (highest-degree first). Shown steps: 000, 050, 100, 150, 200, and the final state (last).}
	\label{fig:gevol-max}
\end{figure*}

\begin{figure*}[h!]
	\centering
	\setlength{\tabcolsep}{2pt}
	\renewcommand{\arraystretch}{1.0}
	\begin{tabular}{cccc}
		\includegraphics[width=0.24\linewidth]{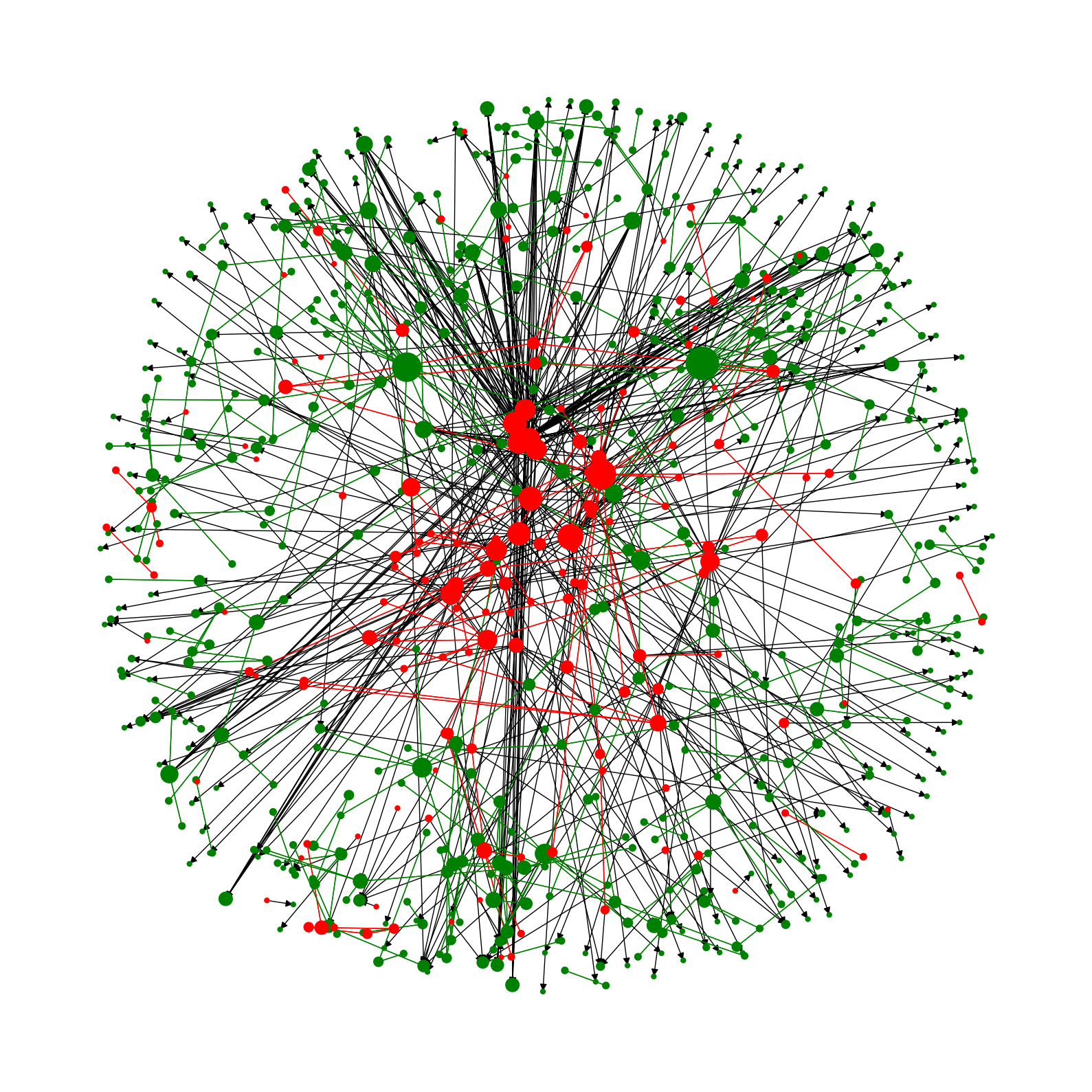} &
		\includegraphics[width=0.24\linewidth]{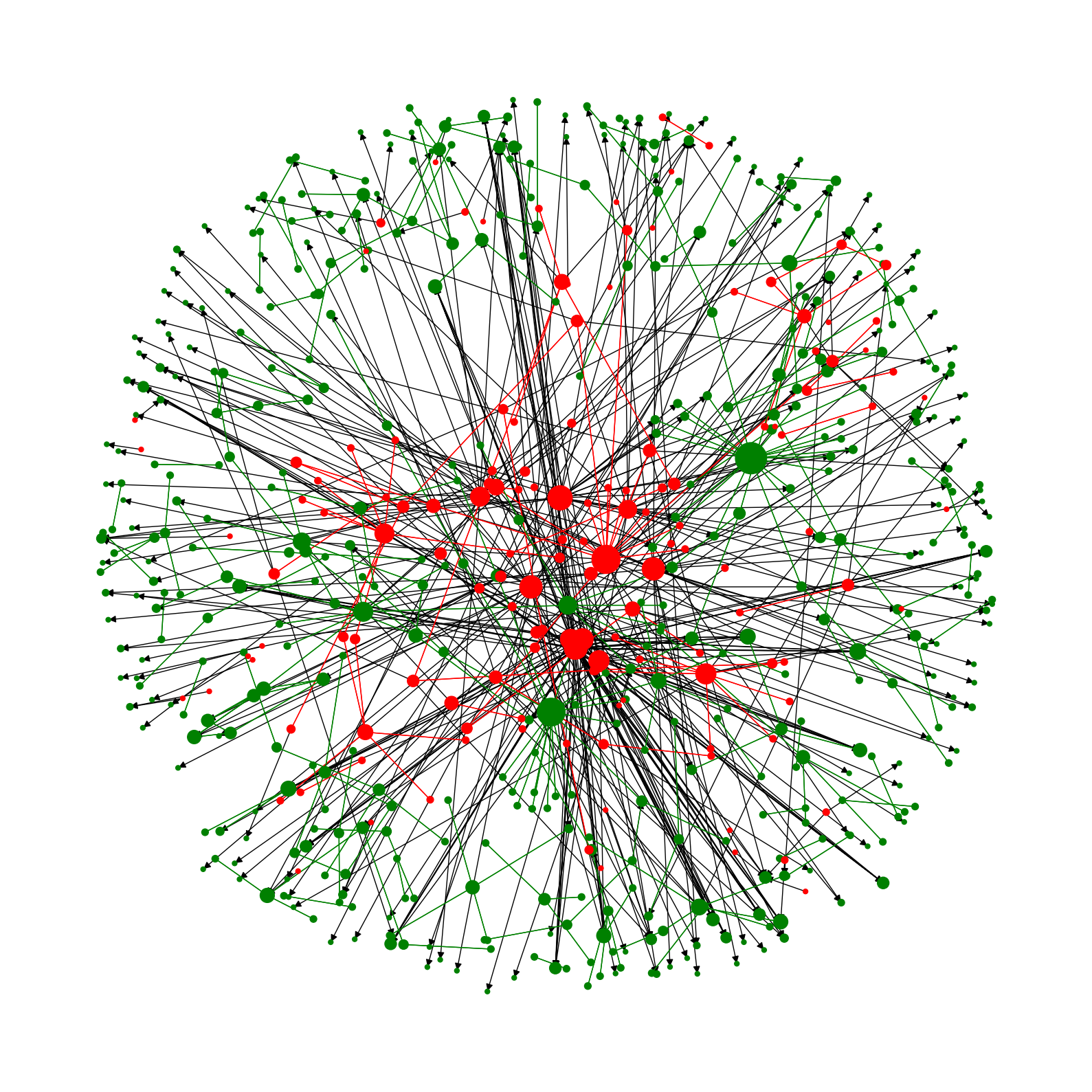} &
		\includegraphics[width=0.24\linewidth]{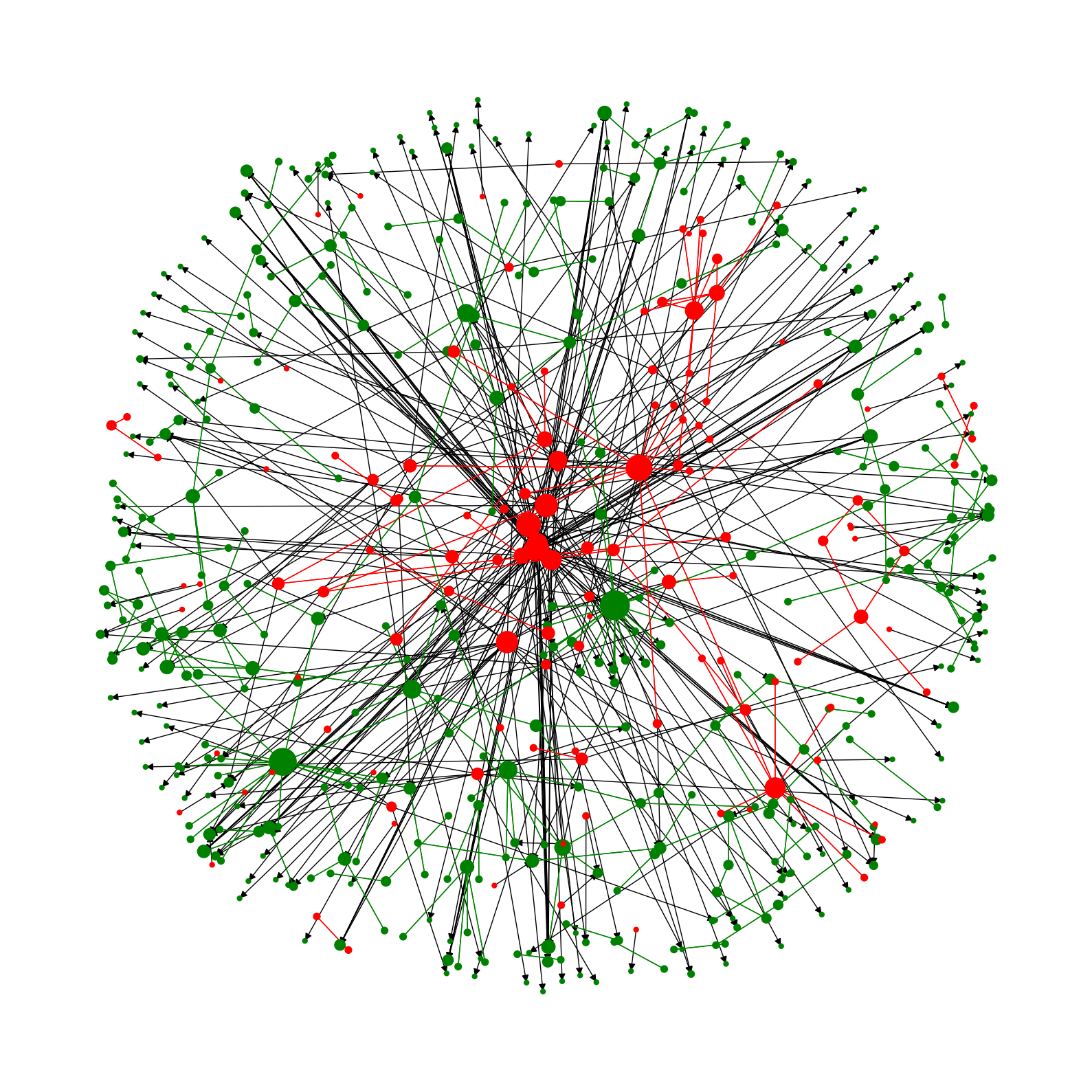} &
		\includegraphics[width=0.24\linewidth]{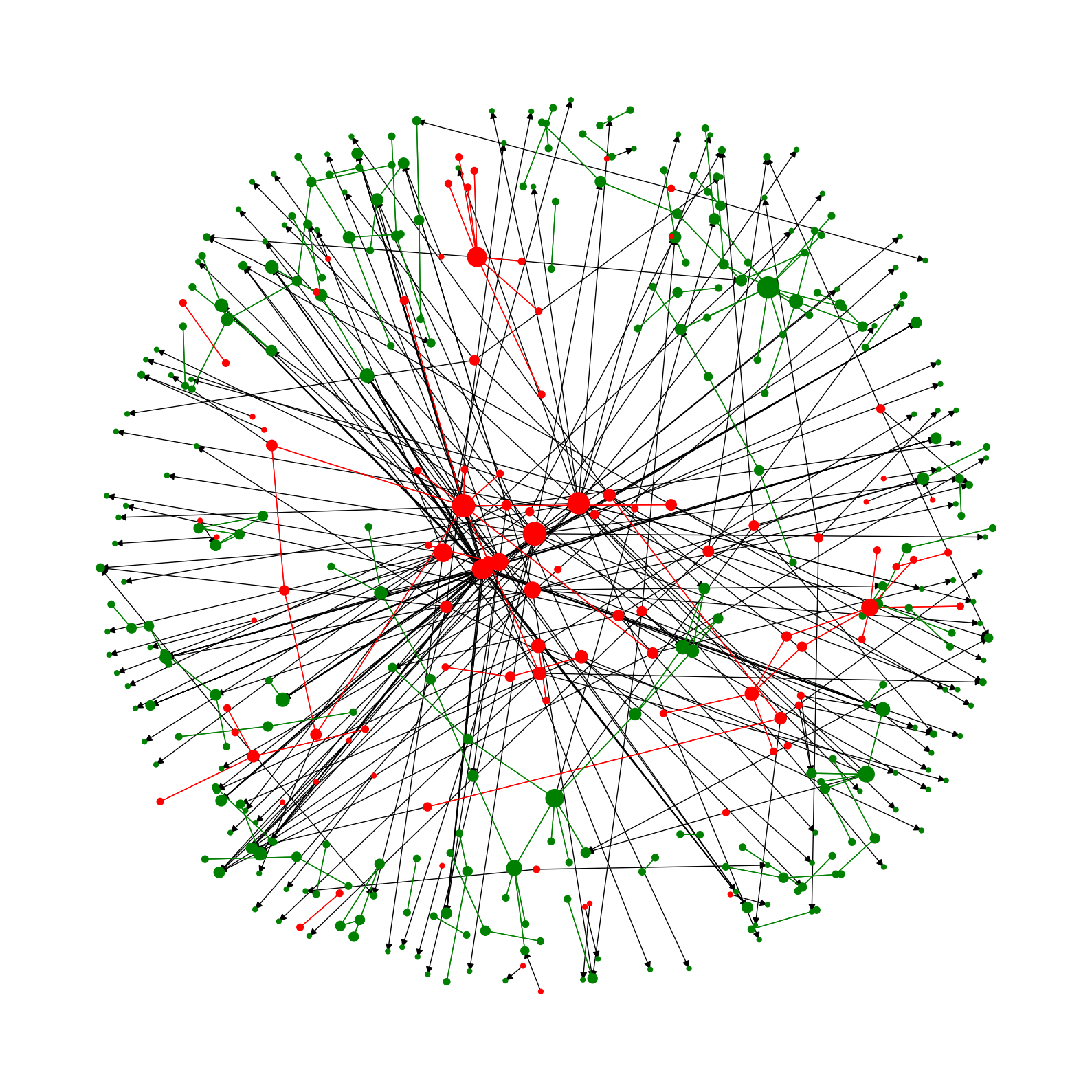} \\
		{\footnotesize step 000} & {\footnotesize step 050} & {\footnotesize step 100} & {\footnotesize step 200} \\
		\includegraphics[width=0.24\linewidth]{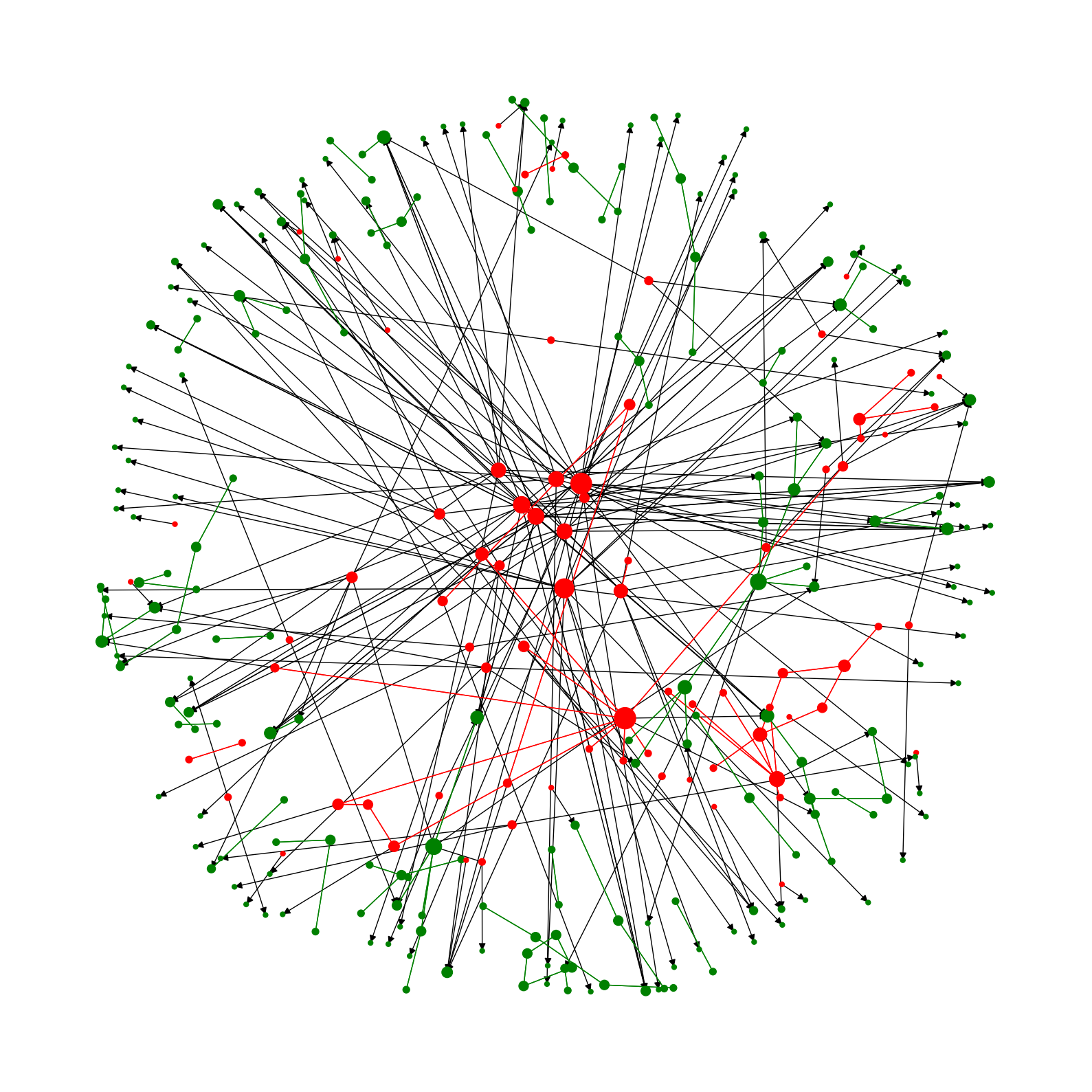} &
		\includegraphics[width=0.24\linewidth]{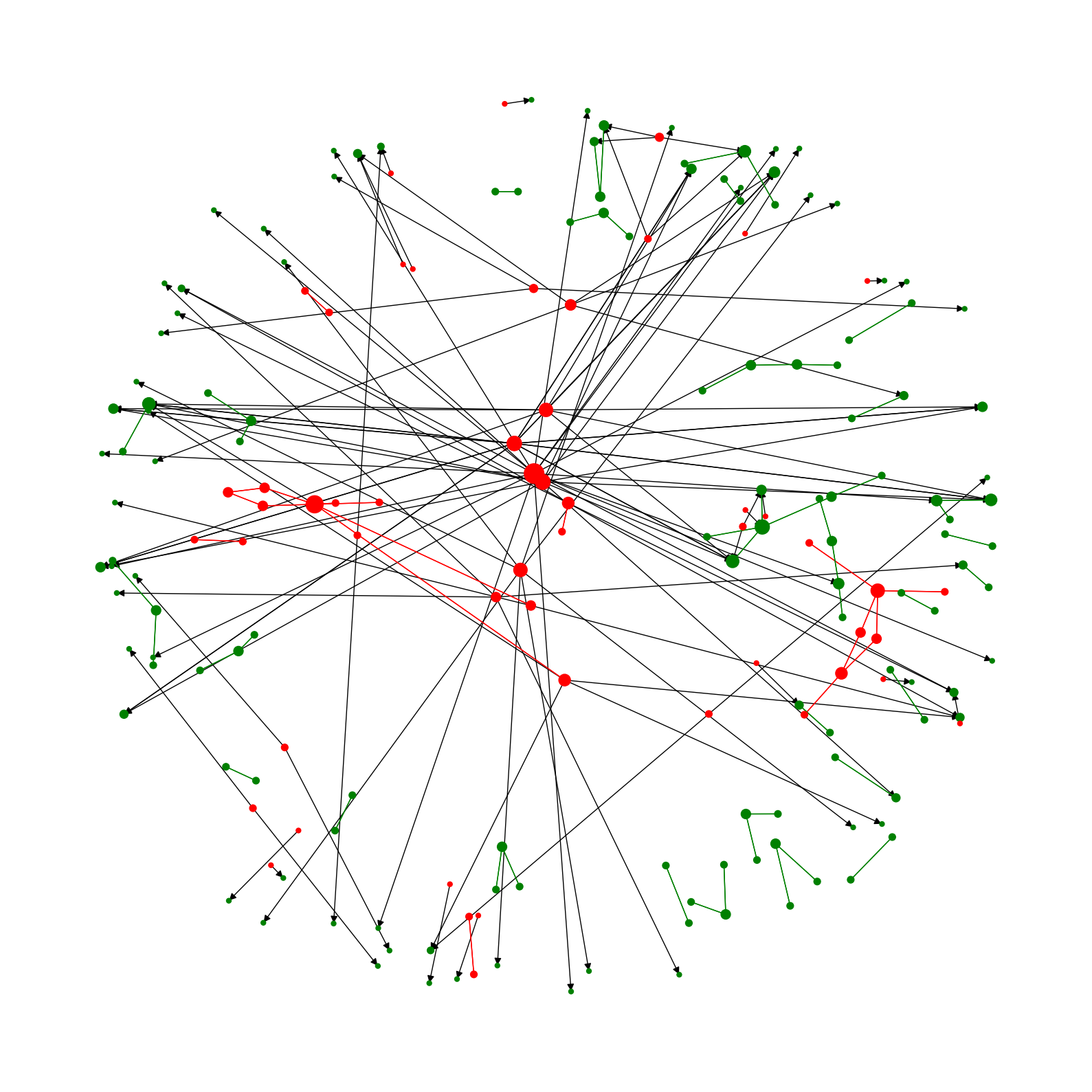} &
		\includegraphics[width=0.24\linewidth]{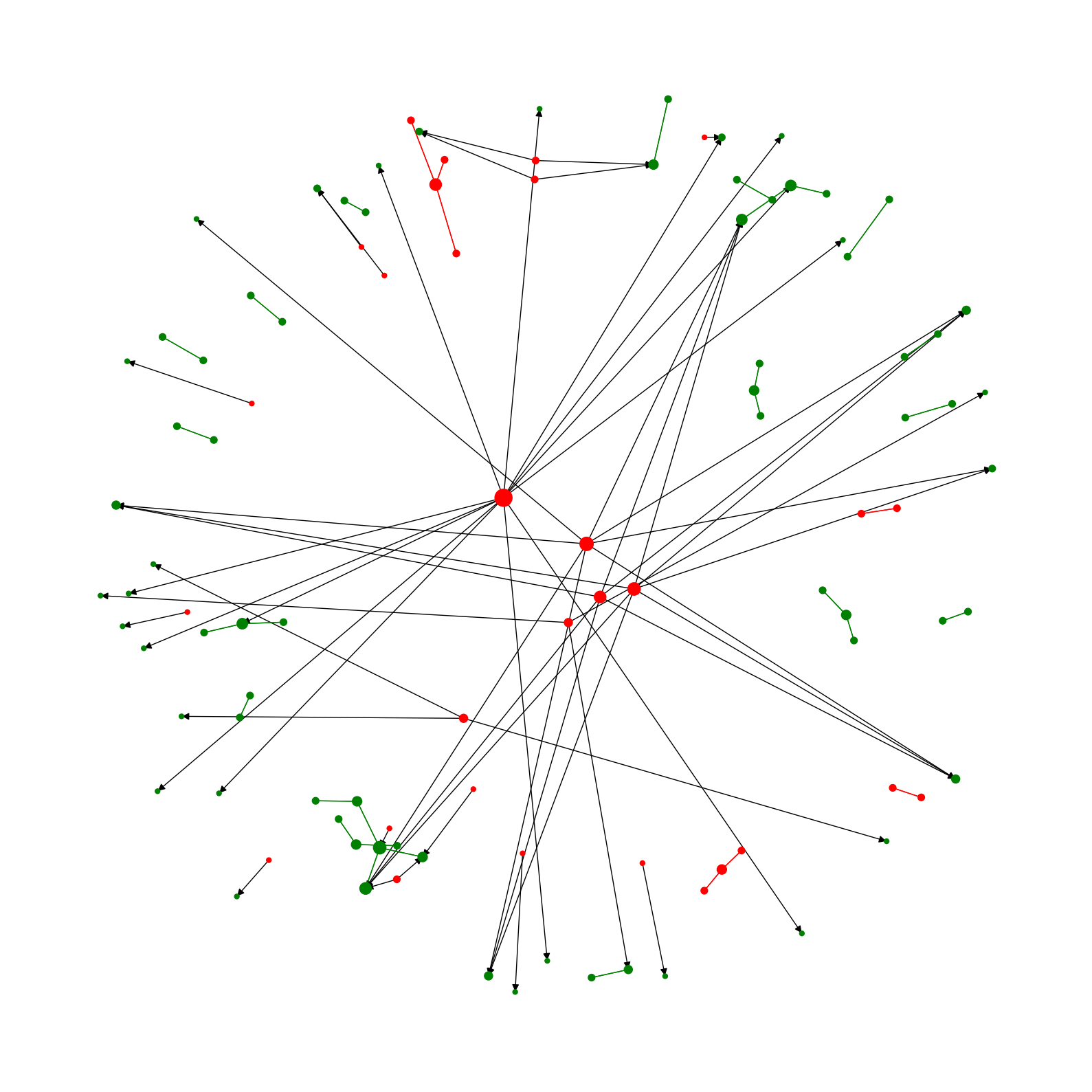} &
		\includegraphics[width=0.24\linewidth]{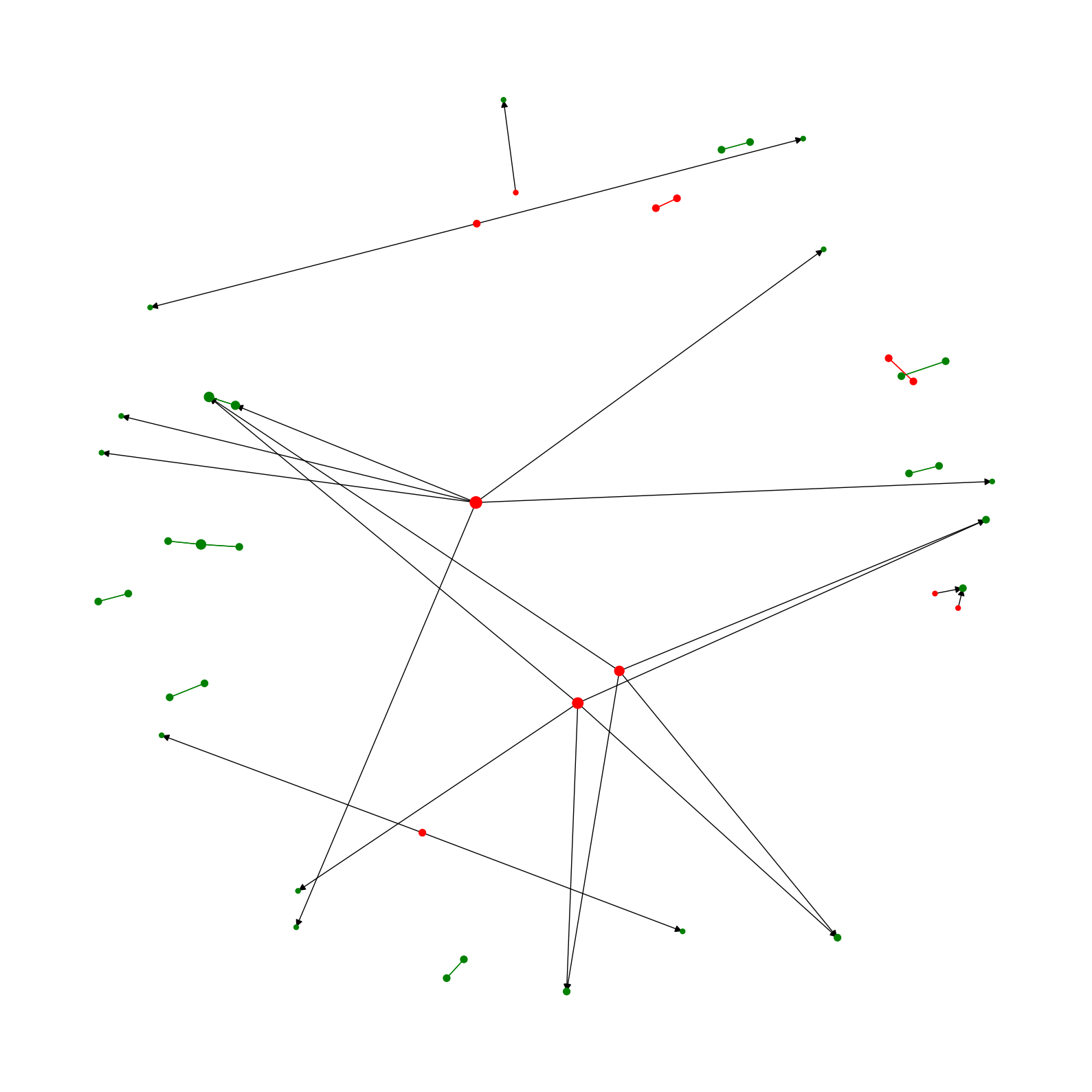}
		\\
		{\footnotesize step 300} & {\footnotesize step 400} & {\footnotesize step 500} & {\footnotesize step 600} \\
	\end{tabular}
	
	\caption{(Colour online) Network evolution snapshots for \textsl{Homo sapiens} under random node removal. Shown steps: 000, 050, 100, 200, 300, 400, 500, 600.}
	\label{fig:gevol-rnd}
\end{figure*}

To illustrate the structural changes during this process, we present in figures~\ref{fig:gevol-max} and \ref{fig:gevol-rnd} several representative snapshots of the network at different stages of node removal.
Additionally, two videos illustrating the network evolution during the node removal are available in the repository referenced in the Data Availability section.


The analysis of network density dynamics during progressive node removal reveals different patterns depending on the removal strategy. When the most influential nodes are removed, the density decreases almost linearly, indicating a uniform degradation of the network structure. 
By contrast, under random node removal (averaged over multiple realizations), the density remains nearly constant over a large fraction of the removal process, reflecting the robustness of the network to random failures. At later stages, however, the density exhibits a sharp increase. This effect arises from the rapid decrease in the number of the remaining nodes, which dominates over the loss of edges and leads to an apparent densification of the residual network. In individual realizations, this process displays pronounced fluctuations, including transient increases and decreases in density, which can give the impression of intermediate fragmentation stages. However, these fluctuations are smoothed out upon averaging, revealing the underlying trend of late-stage densification.
These dependencies are visualized in table~\ref{tab:density}, which presents the corresponding density evolution graphs for different node removal scenarios.

The dynamics of the degree assortativity coefficient during gradual node removal reveal distinctions among the networks under study. In the case of removing the most influential nodes in the {\sl Homo sapiens} network, the initial assortativity value is close to zero. After the removal of nine nodes, the coefficient reaches a peak, followed by a sharp drop to small negative values.
Subsequently, a gradual increase is observed, although it does not reach the initial peak. Just before the structural collapse of the graph, the assortativity again approaches zero. In the {\sl Gallus gallus} network, disassortativity is observed at all stages. For {\sl Mus musculus}, the assortativity values also remain negative for the  most part of the process but become positive just before the network's final breakdown.

\begin{table}[h] 
	\caption{Evolution of network density during step-by-step node removal. The vertical axis shows the current network density $D$, while the horizontal axis represents the number of the removed nodes. Separate plots are shown for two scenarios: removal of the most influential hubs and random node deletion. The latter corresponds to the values averaged over 100 independent realizations.}
	\label{tab:density}
	\scriptsize
	\centering
	
	\begin{tabular}{lcc}
		\hline\noalign{\smallskip} 
		\footnotesize{\bf Species} & 
		\footnotesize{\bf Hubs removed} &
		\footnotesize{\bf Random}\\
		\noalign{\smallskip}\hline\noalign{\smallskip}
		\raisebox{6em}{\sl Homo sapiens} & 
		\includegraphics[scale=0.36]{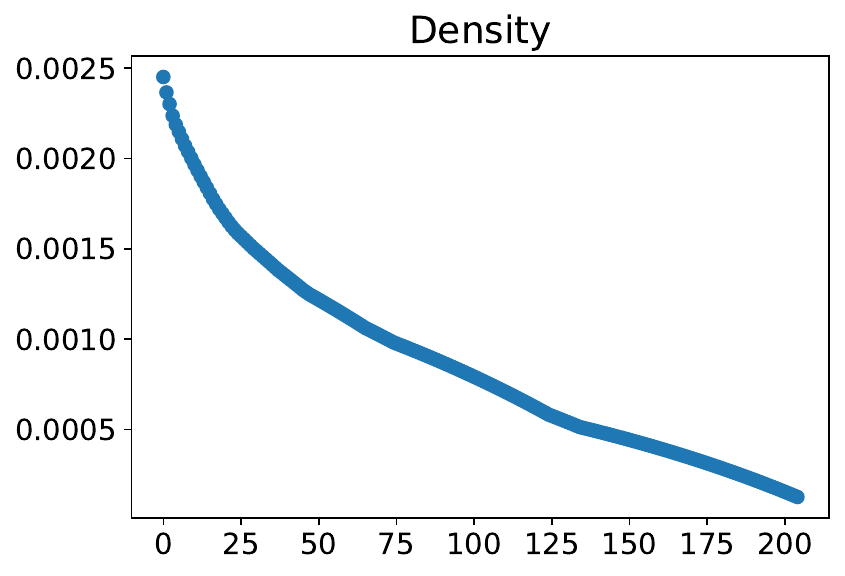} &
		\includegraphics[scale=0.36]{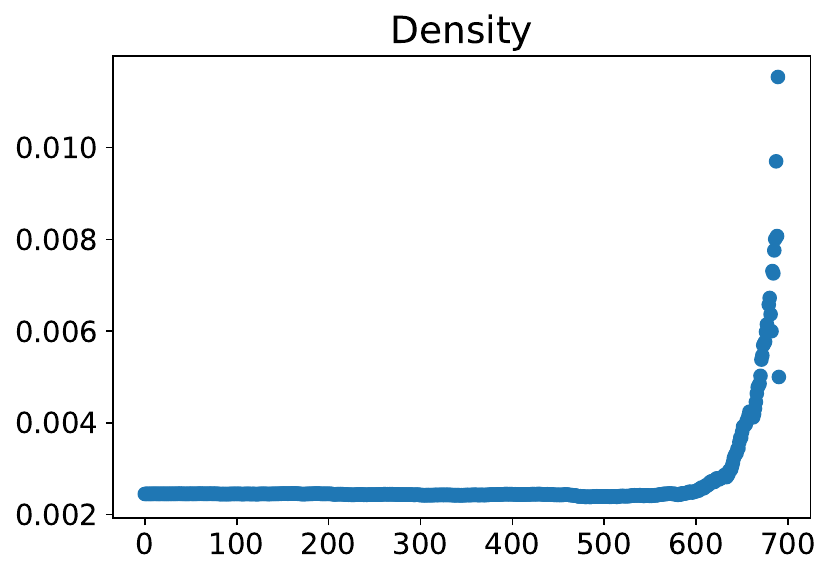}\\
		\raisebox{6em}{\sl Gallus gallus} &
		\includegraphics[scale=0.36]{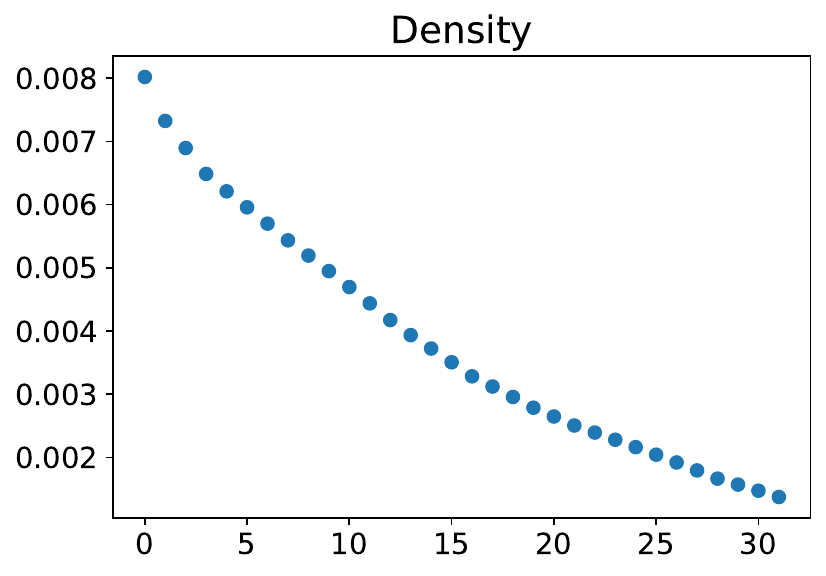} &
		\includegraphics[scale=0.36]{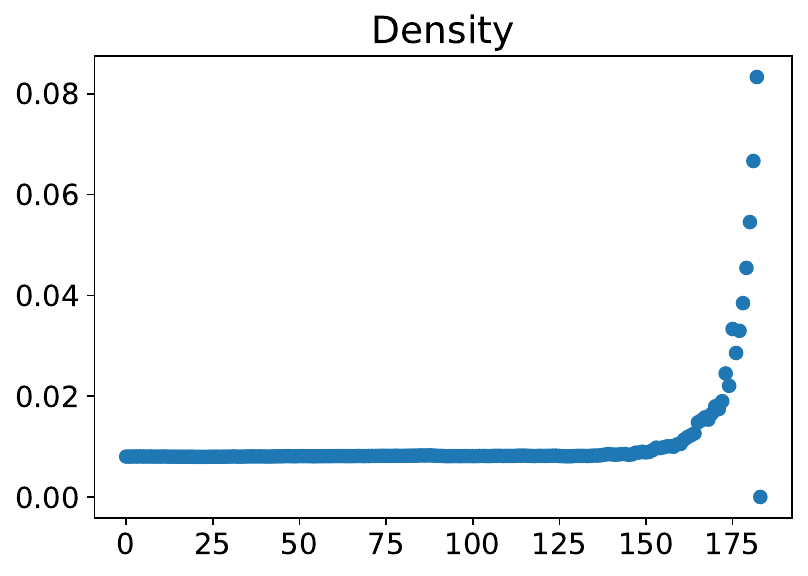}\\
		\raisebox{6em}{\sl Mus musculus} &
		\includegraphics[scale=0.36]{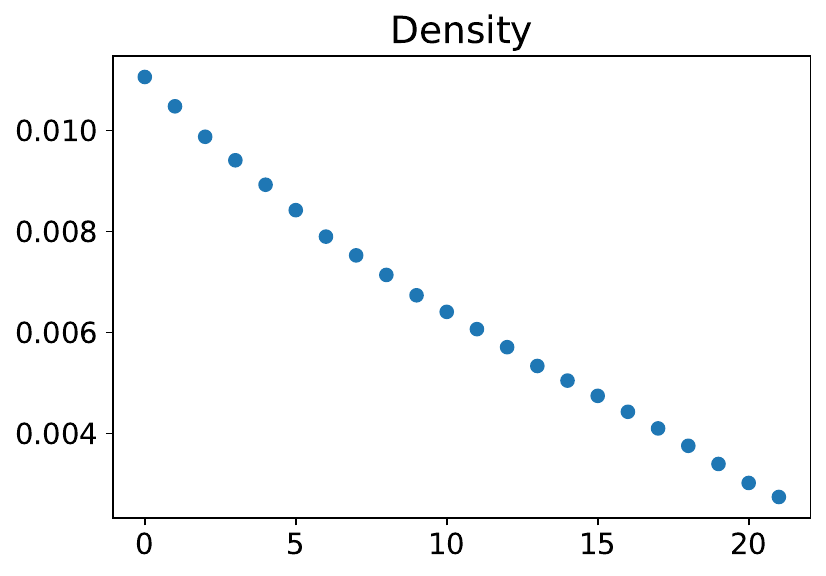} &
		\includegraphics[scale=0.36]{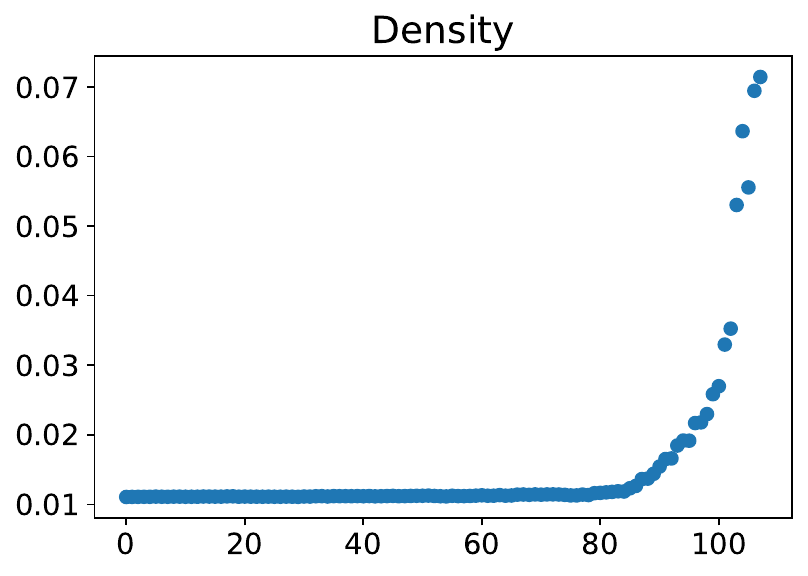}\\
		\noalign{\smallskip} \hline
	\end{tabular}
\end{table}

During random node removal (averaged over multiple realizations), the assortativity coefficient exhibits a smoother behavior. In the {\sl Homo sapiens} network, it remains a small positive value over a large portion of the process, followed by a gradual decrease and a pronounced drop to negative values at late stages. In the case of {\sl Gallus gallus}, assortativity remains negative throughout, with only minor variations before a sharper decline near the end. For {\sl Mus musculus}, disassortativity is also preserved across the entire interval, with a slight increase in the intermediate regime followed by a decrease at later stages. These trends indicate that, while targeted removal induces pronounced structural rearrangements, random removal leads to more gradual and averaged changes in degree correlations.
The corresponding changes in the assortativity coefficient are visualized in table~\ref{tab:assortativity}.

\begin{table}[h] 
	    \caption{Change in degree assortativity coefficient during progressive node removal. Both hub-targeted and random deletion strategies are presented for comparison.}
	\label{tab:assortativity}
    \scriptsize
    \centering

    \begin{tabular}{lcc}
    \hline\noalign{\smallskip}
        \footnotesize{\bf Species} & 
        \footnotesize{\bf Hubs removed} & 
        \footnotesize{\bf Random}\\
    \noalign{\smallskip}\hline\noalign{\smallskip}
        \raisebox{6em}{\sl Homo sapiens} & 
        \includegraphics[scale=0.36]{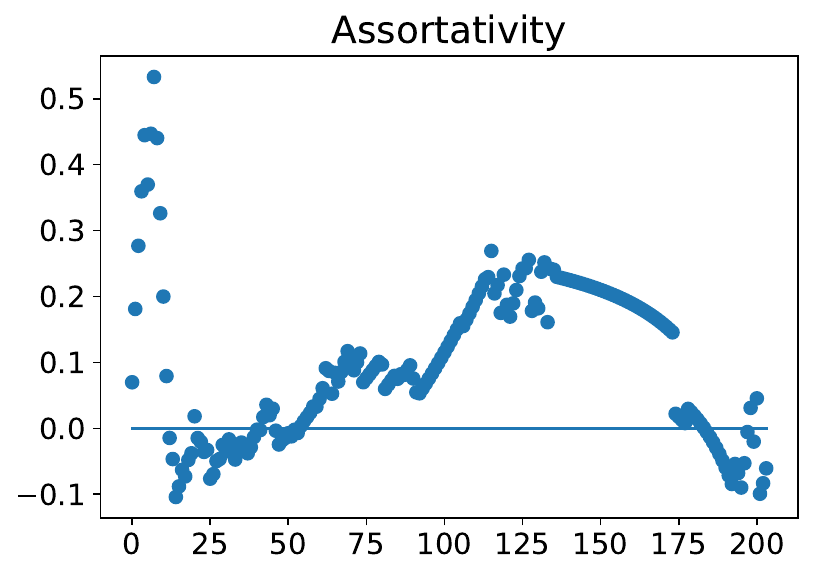} &
        \includegraphics[scale=0.36]{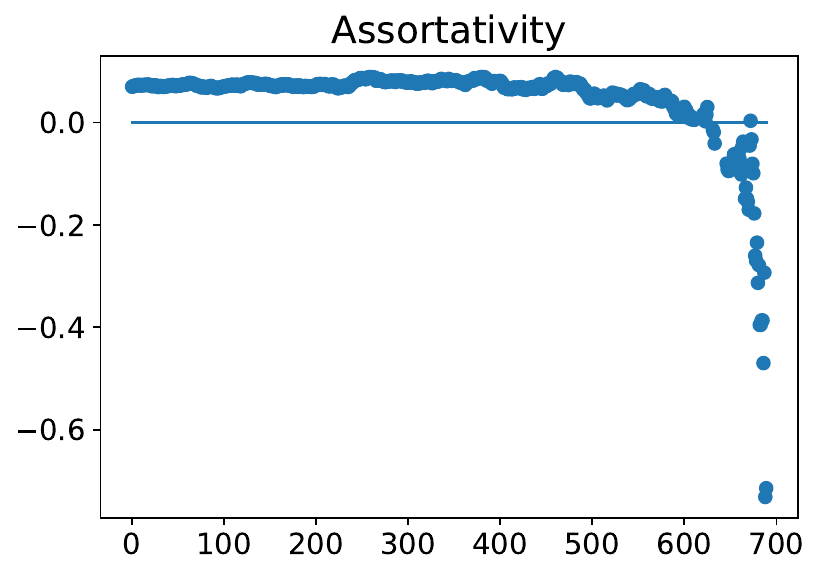}\\
        \raisebox{6em}{\sl Gallus gallus} &
        \includegraphics[scale=0.36]{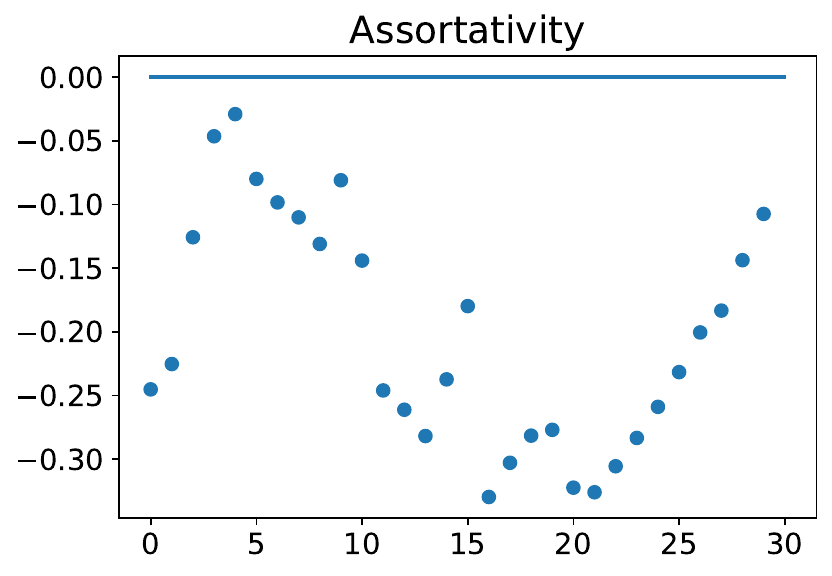} &
        \includegraphics[scale=0.36]{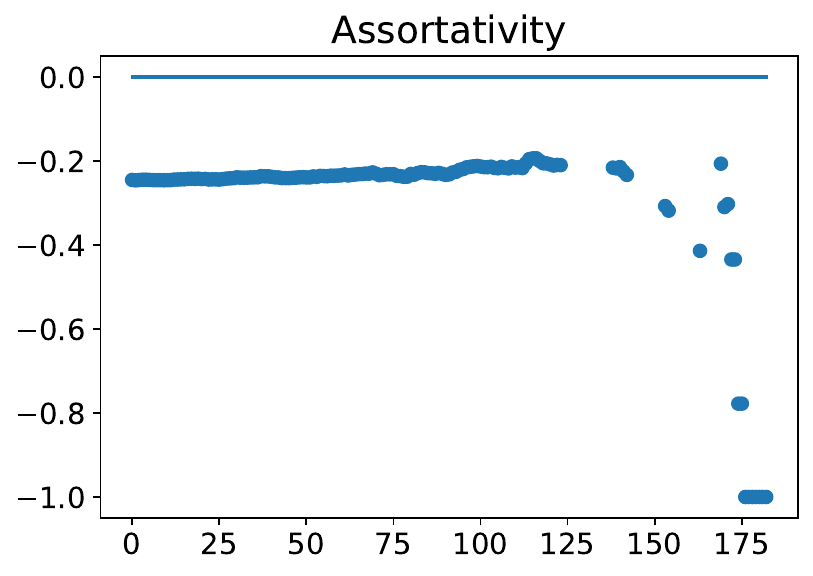}\\
        \raisebox{6em}{\sl Mus musculus} &
        \includegraphics[scale=0.36]{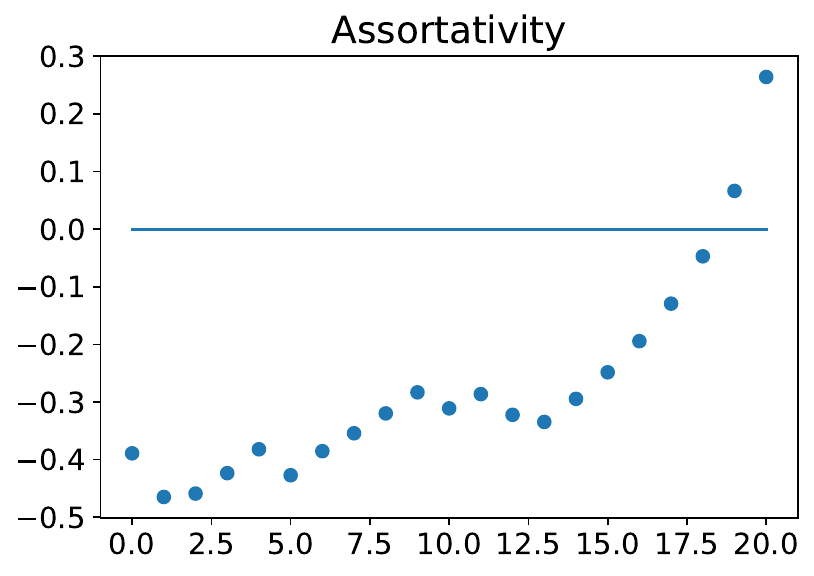} &
        \includegraphics[scale=0.36]{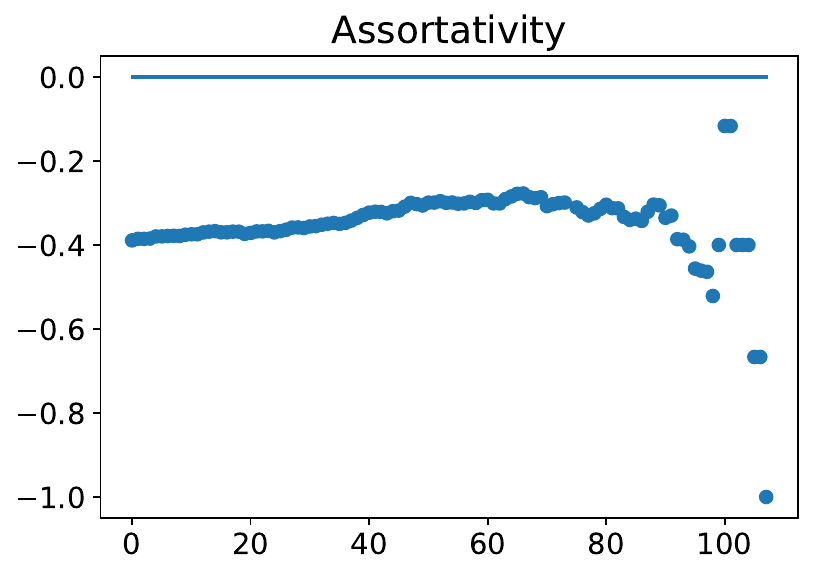}\\
        \noalign{\smallskip}\hline
        \end{tabular}
\end{table}

The dynamics of graph temperature during the node removal demonstrate stability throughout most of the process, indicating the resilience of thermodynamic properties of the network to gradual size reduction. When the most influential nodes are removed from the {\sl Homo sapiens} network, the temperature remains almost unchanged until the final stages of network collapse. Only near the critical point there is observed a sharp temperature drop, which may indicate a breakdown of global connectivity. After reaching a minimum, a slight increase is possible, probably associated with the reorganization of the residual structure before final disintegration. A similar pattern is observed for {\sl Mus musculus}, although the temperature drop is less pronounced and the graph displays a characteristic ``dip'', that may correspond to a specific stage of structural reconfiguration. In the case of {\sl Gallus gallus}, the temperature remains nearly constant throughout the entire process, which may indicate a lower sensitivity of  thermodynamic characteristics of the network to structural changes.

In the case of random node removal (averaged over multiple realizations), a similar overall stability is observed, with smoother trends and reduced variability. For {\sl Homo sapiens}, the temperature remains nearly constant over the most part of the process, followed by a gradual decrease and a more pronounced drop at late stages. For {\sl Gallus gallus}, the temperature exhibits a slow, monotonous decline with minor deviations, without significant abrupt changes. In {\sl Mus musculus}, a gradual decrease is also observed, with a more noticeable drop near the final stages of network degradation. Overall, the averaged behavior indicates that random removal preserves a thermodynamic stability over a wide range of the node removal, with significant changes occurring primarily close to the network collapse.

The temperature dynamics during the node removal are presented in table~\ref{tab:temp}, which displays the corresponding graphs for each of the analyzed species.

\begin{table}[ht] 
	    \caption{Graph temperature as a function of the number of the removed nodes. Differences between targeted and random removal strategies are visualized.}
	\label{tab:temp}
    \scriptsize
    \centering

    \begin{tabular}{lcc}
    \hline\noalign{\smallskip}
        \footnotesize{\bf Species} & \footnotesize{\bf Hubs removed} & \footnotesize{\bf Random}\\
    \noalign{\smallskip}\hline\noalign{\smallskip}
        \raisebox{6em}{\sl Homo sapiens} & \
        \includegraphics[scale=0.36]{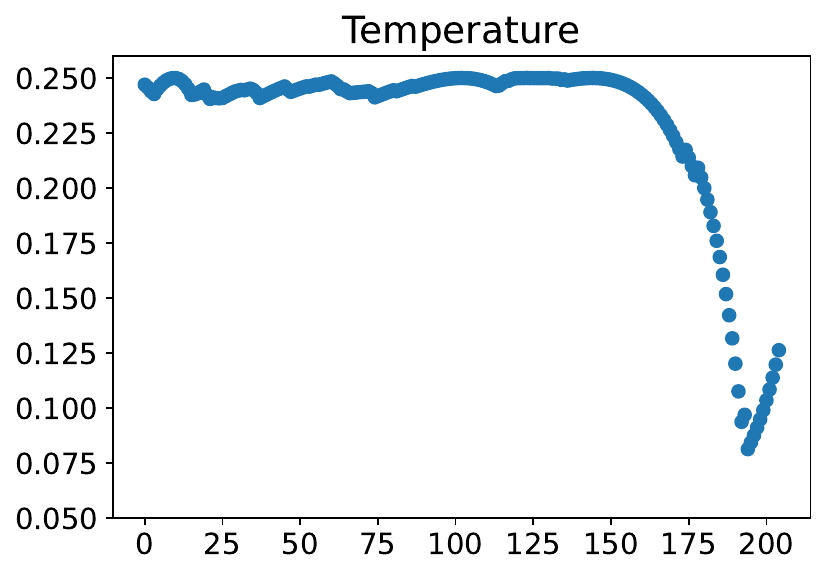} &
        \includegraphics[scale=0.36]{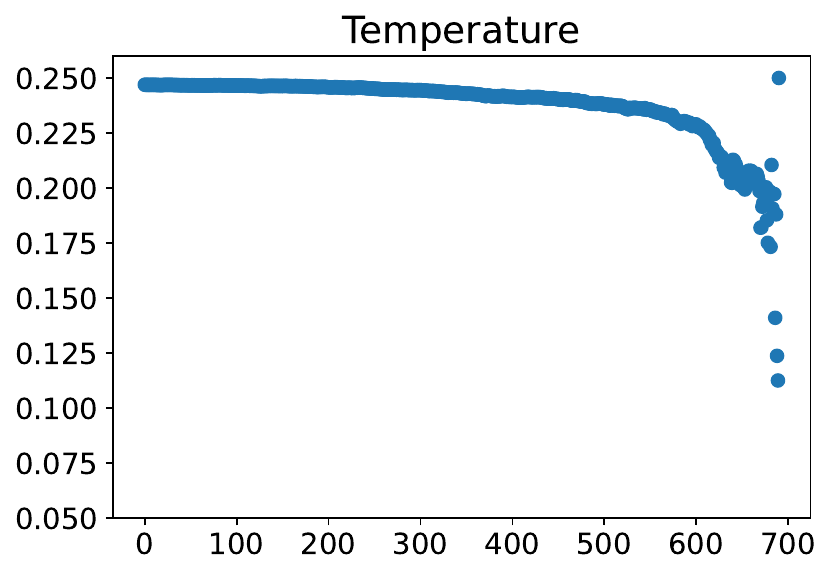}\\
        \raisebox{6em}{\sl Gallus gallus} &
        \includegraphics[scale=0.36]{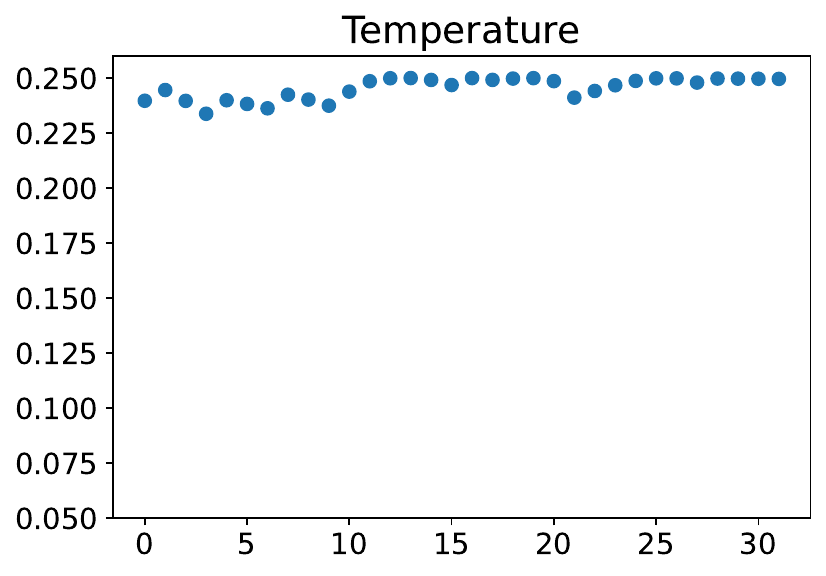} &
        \includegraphics[scale=0.36]{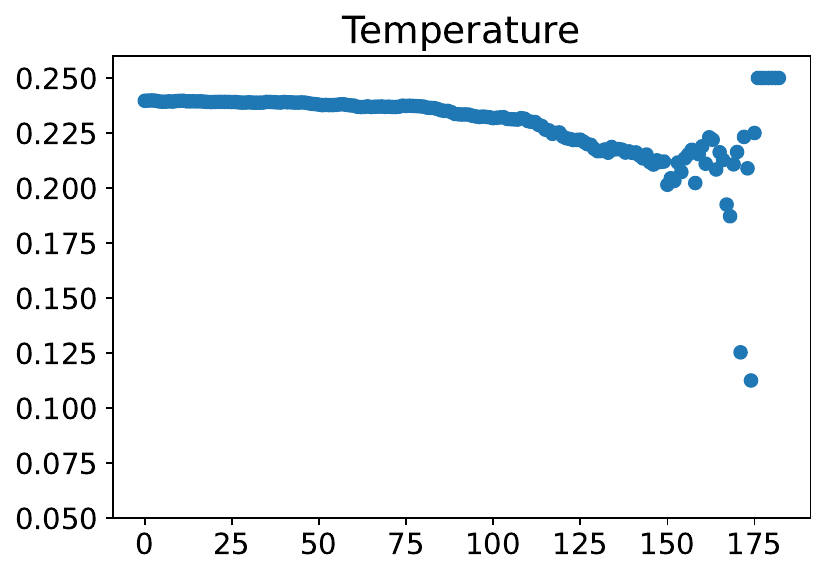}\\
        \raisebox{6em}{\sl Mus musculus} &
        \includegraphics[scale=0.36]{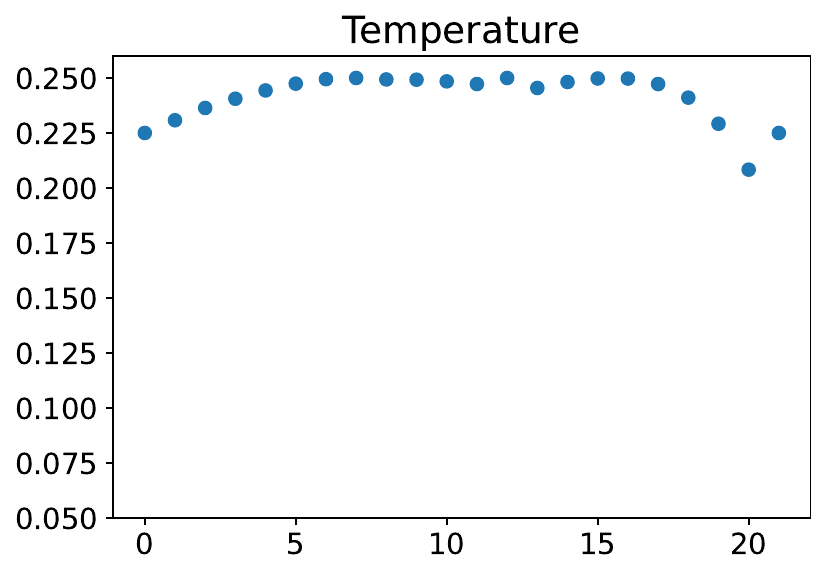} &
        \includegraphics[scale=0.36]{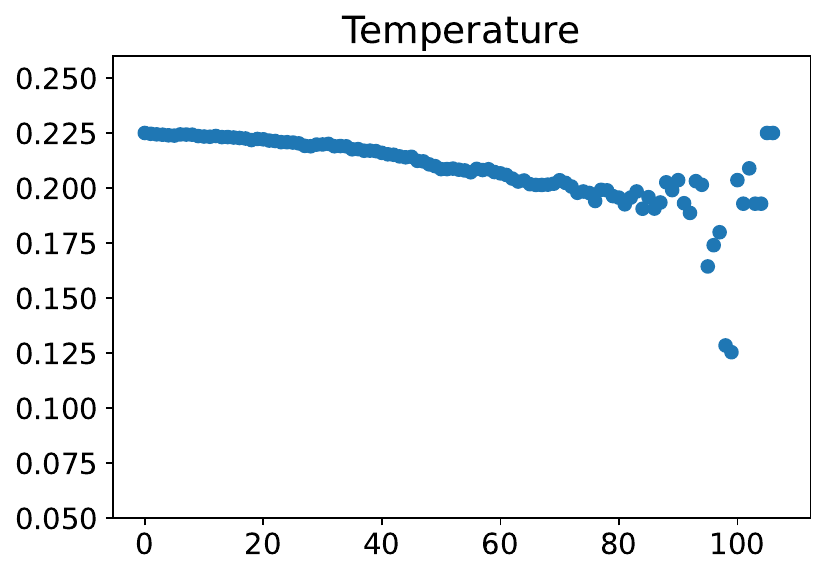}\\
        \noalign{\smallskip}\hline
        \end{tabular}
\end{table}

\subsection{Susceptibility}

To analyze the structural properties of the graph, one can employ the concept of magnetization, which originates from statistical physics and phase transition theory. It characterizes the balance between two types of nodes in the network and accounts for their interrelations via the adjacency matrix.

Magnetization is defined as:
\begin{align}\label{eq:M-def}
    M = \frac{1}{N}\sum_{i, j} A_{ij}s_j,
\end{align}
\noindent where $A$ is the weighted adjacency matrix of the graph, $N$ is the number of nodes, and $s_j$ is a spin variable that characterizes the node type \cite{barrat2008dynamical}. For further analysis, let us assume that the nodes representing viruses are assigned $s_j = +1$, while those representing hosts are assigned $s_j = -1$.

Due to fixed spin assignments, the magnetization defined in equation~(\ref{eq:M-def}) reduces  to a weighted imbalance between different classes of interactions (virus--virus, host--host, and cross interactions). Therefore, it serves as a structural indicator of interaction asymmetry rather than a thermodynamic order parameter in the traditional sense.

Here, the magnetization can be written as the average of the node-level contributions, 
\begin{align}
M = \frac{1}{N}\sum_i x_i,\qquad \text{where}\quad x_i = \sum_j A_{ij}s_j.
\end{align}

One of the important parameters characterizing the network is its susceptibility. It measures how sensitive the system is to the changes in the structure of the graph, analogous to how susceptibility in physical systems describes the response to external perturbations \cite{castellano2009statistical}. 
It is defined as:
\begin{align}\label{eq:chi-def}
    \chi = \frac{1}{T}\left(\langle x_i^2\rangle - \langle x_i\rangle^2\right),
\end{align}
\noindent where $T$ is the graph temperature. Accordingly, the susceptibility is defined as the variance of these node-level contributions.

In contrast to the conventional definition of susceptibility in statistical mechanics, the averaging in equation~(\ref{eq:chi-def}) is performed over the  node-level quantities within a single network configuration, rather than over an ensemble of configurations. Therefore, $\chi$ quantifies the dispersion of local interaction imbalances and should be interpreted as a structural heterogeneity measure rather than a thermodynamic fluctuation.

High values of susceptibility indicate an increased structural heterogeneity and may signal the points of pronounced structural reorganization, while low values suggest a more homogeneous network configuration with a stable structure less prone to changes.

The dynamics of susceptibility can serve as an indicator of structural changes in the graph during the gradual node removal, allowing the detection of critical points at which the network loses its integrity or alters its fundamental properties.

\begin{figure}[h!]
    \centering
    \includegraphics[scale=0.5]{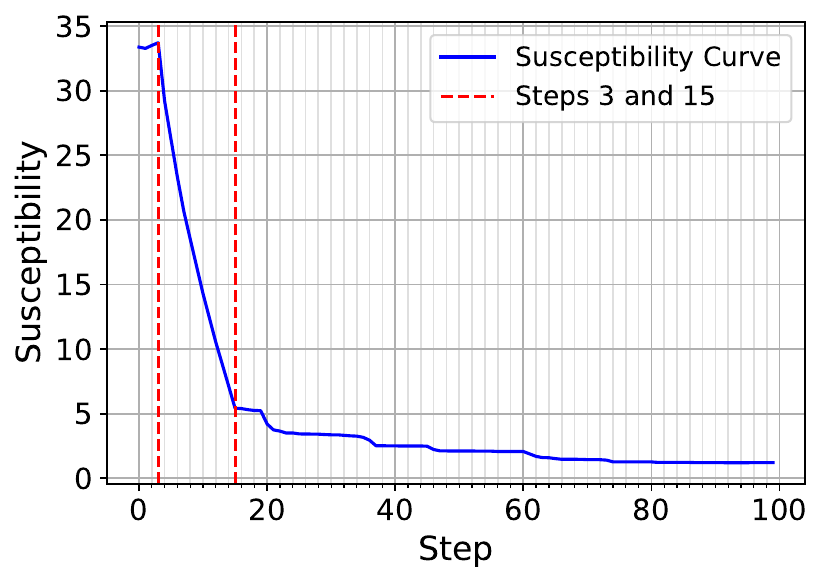}
    \vspace*{-1em}
    \caption{(Colour online) Susceptibility as a function of the number of the removed nodes in the {\sl Homo sapiens} network. The plot shows two inflection points, indicating structural transitions during progressive node removal.}
    \label{fig:homo-chi}
\end{figure}

From the dependence of susceptibility on the number of the removed nodes for {\sl Homo sapiens}, shown in figure~\ref{fig:homo-chi}, two points can be identified where the slope of the curve changes noticeably. Similarly, in {\sl Gallus gallus}, two inflection points are observed: one again at the third step, and the second one located in the range of approximately 12--18 steps. In the case of {\sl Mus musculus}, the analogy is less clear, as the network becomes sparse after a relatively small number of steps due to the limited amount of data. Nevertheless, a sharp drop between the 12th and 13th steps is still clearly visible.

\begin{figure}[ht]
    \centering
    {\sl Gallus gallus} \hspace*{5cm} {\sl Mus musculus}\\
    \includegraphics[scale=0.5]{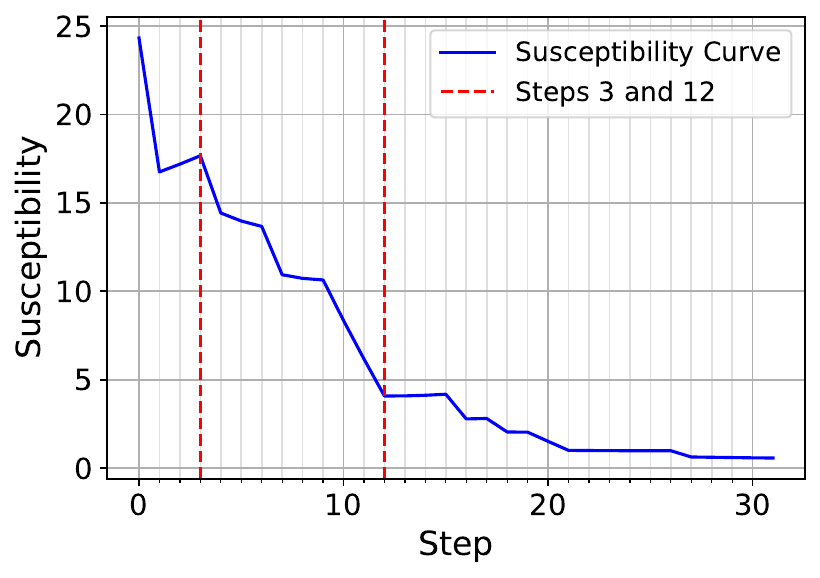}
    \quad
    \includegraphics[scale=0.5]{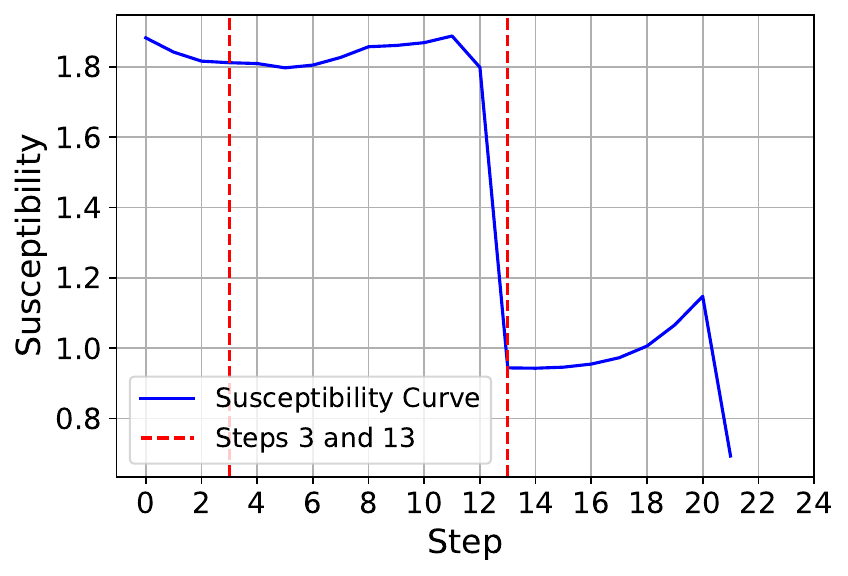}
    \caption{(Colour online) Susceptibility dynamics for {\sl Gallus gallus} and {\sl Mus musculus}. Characteristic slope changes indicate potential points of structural fragility in the corresponding networks.}
    \label{fig:others-chi}
\end{figure}

Certain characteristic patterns can also be observed in assortativity and betweenness at these points, as shown in the plots in table~\ref{tab:betw-s}. For all studied species, betweenness begins to decline sharply after the second point. The assortativity parameter exhibits a behavioral pattern as well: prior to the first point, an increase is generally observed; between the first and second points, there is a notable decline --- except in the case of {\sl Mus musculus};  after the second point, a somewhat pronounced increase is evident.

\begin{table}[h] 
	\caption{Behavior of assortativity and betweenness centrality near susceptibility inflection points. The $Y$-axis on each plot indicates the respective network parameter, while the $X$-axis reflects the progression of the node removal. The highlighted intervals correspond to the regions around critical structural transitions in the network.}
	\label{tab:betw-s}
	\scriptsize
	\centering
	
	\begin{tabular}{lcc}
		\hline\noalign{\smallskip}
		\footnotesize{\bf Species} & 
		\footnotesize{\bf Assortativity} & 
		\footnotesize{\bf Betweenness}\\
		\noalign{\smallskip}\hline\noalign{\smallskip}
		\raisebox{6em}{\sl Homo sapiens} & 
		\includegraphics[scale=0.36]{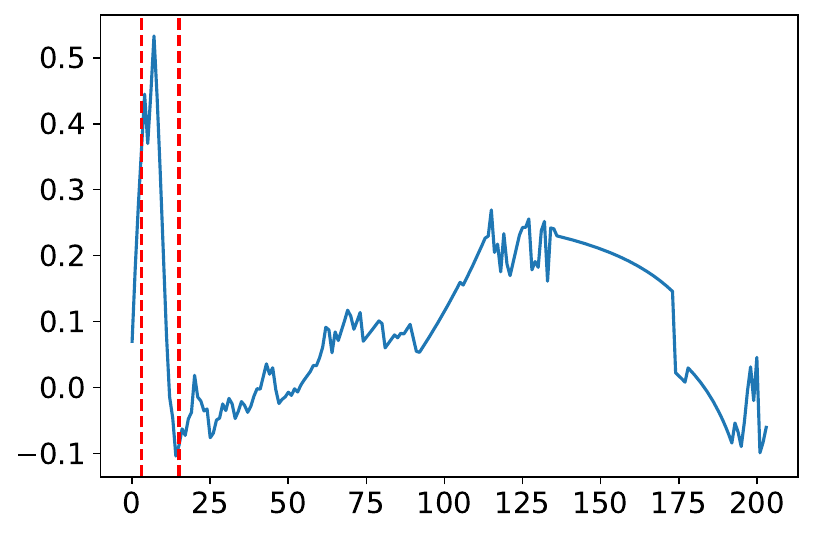} &
		\includegraphics[scale=0.36]{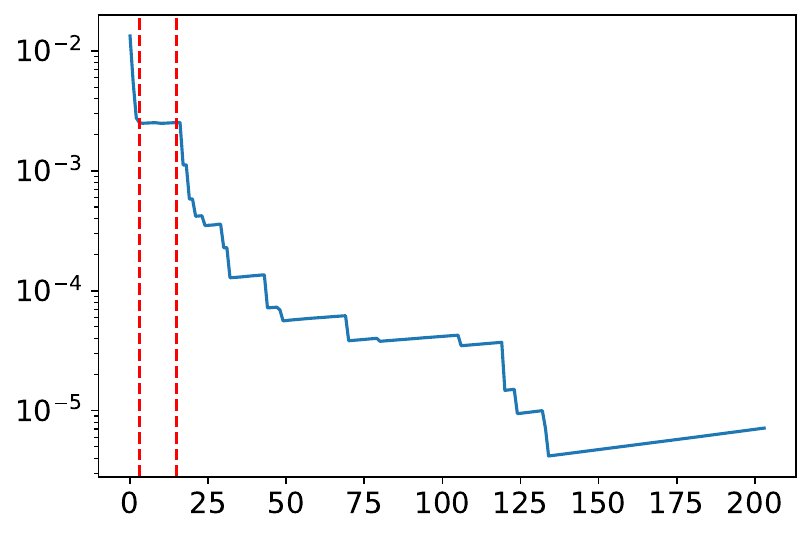}\\
		\raisebox{6em}{\sl Gallus gallus} &
		\includegraphics[scale=0.36]{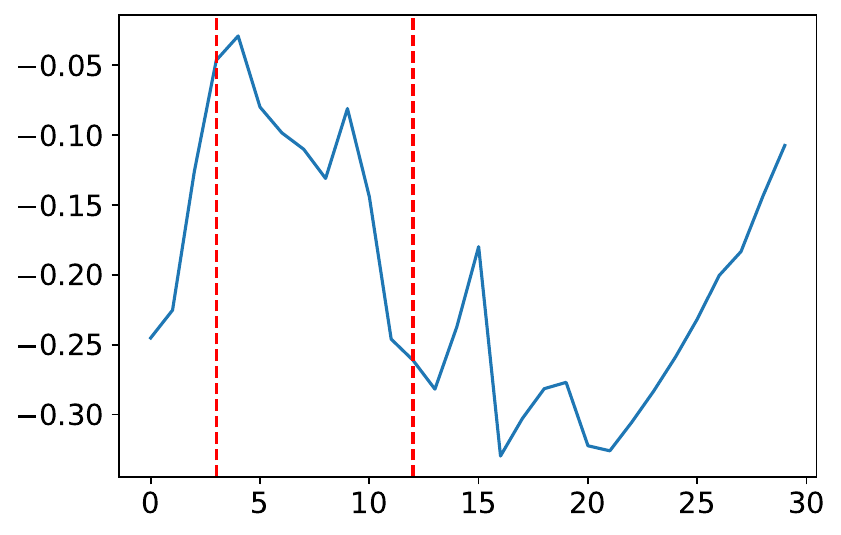} &
		\includegraphics[scale=0.36]{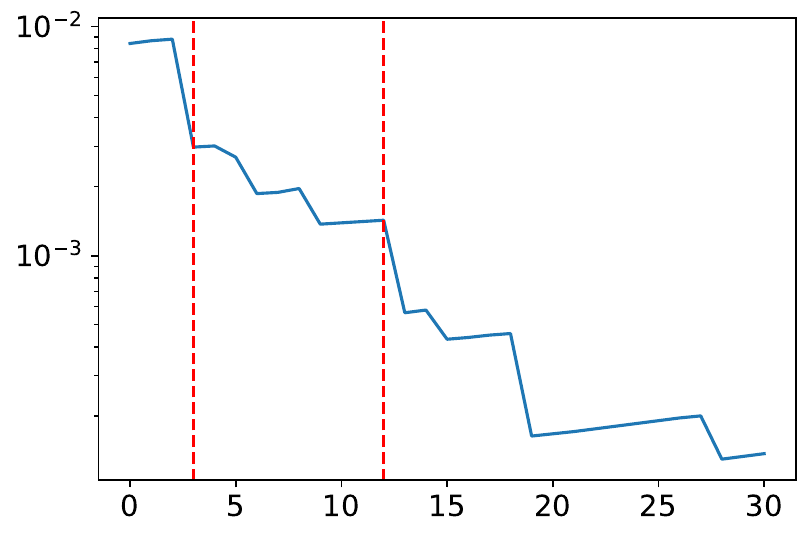}\\
		\raisebox{6em}{\sl Mus musculus} &
		\includegraphics[scale=0.36]{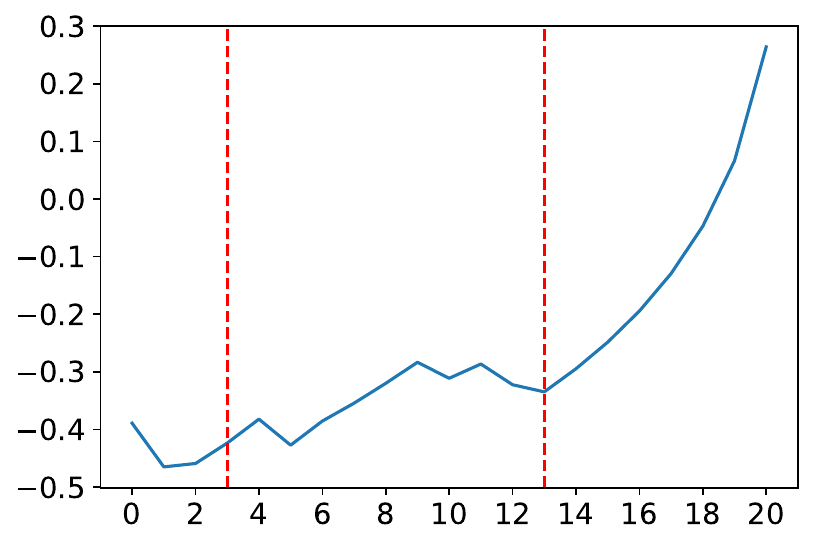} &
		\includegraphics[scale=0.36]{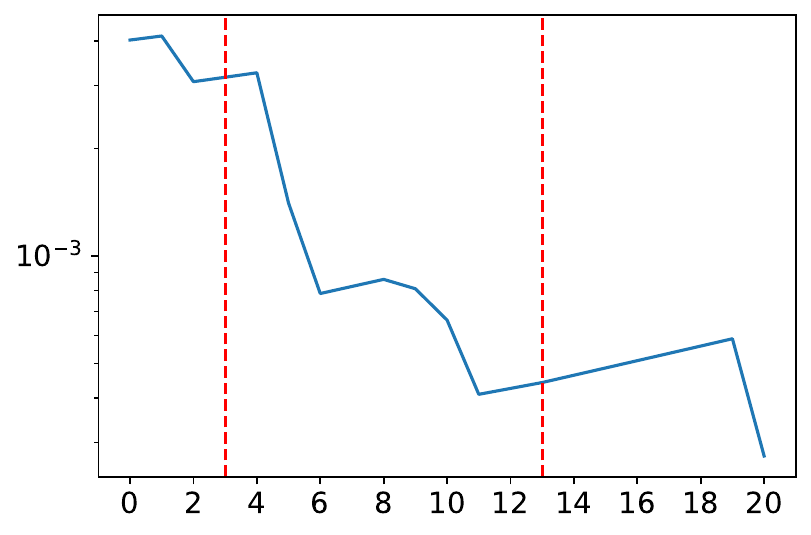}\\
		\noalign{\smallskip}\hline
	\end{tabular}
\end{table}

\subsection{Heat capacity}
\begin{figure}[ht]
\centering
\includegraphics[scale=0.5]{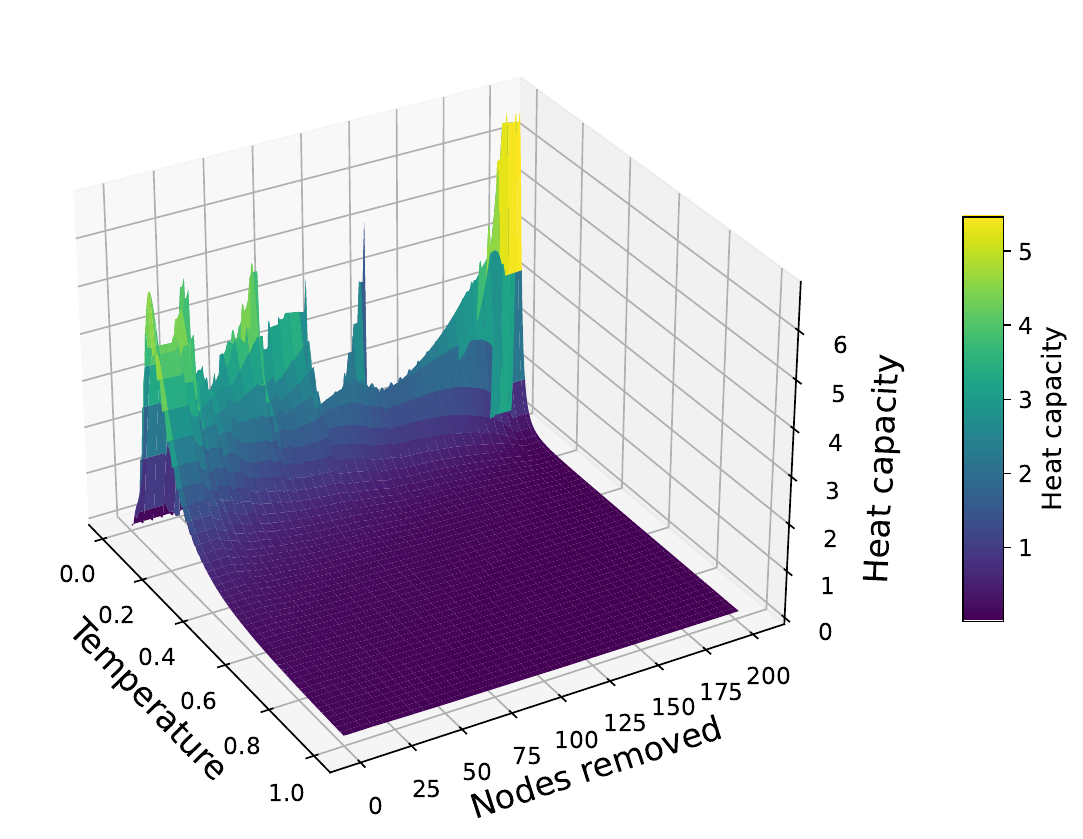}
\caption{(Colour online) Heat capacity \eqref{heat-def} for the \textsl{Homo sapiens} network at targeted node removal.
Note that the temperature here corresponds to the formal thermodynamic temperature $T^* = 1/\beta$.}
\label{fig:heat3D}
\end{figure}

\begin{figure}[h]
	\centering
	\includegraphics[scale=0.5]{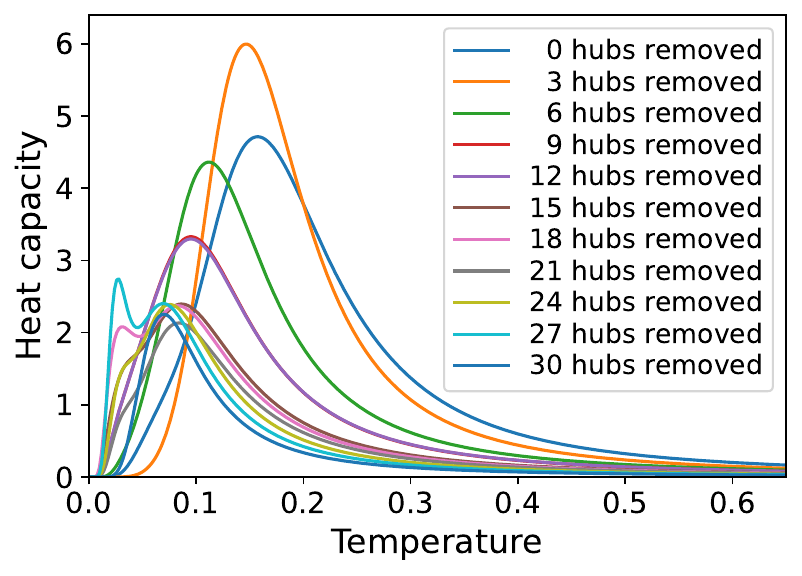}
	\quad
	\includegraphics[scale=0.5]{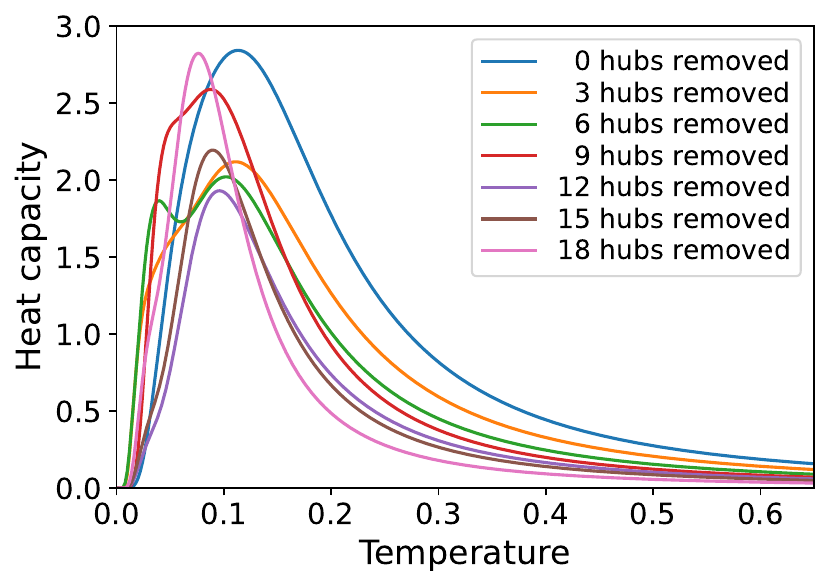}
	\caption{(Colour online) Slices of heat capacity \eqref{heat-def} for the \textsl{Gallus gallus} (left-hand) and \textsl{Mus musculus} (right-hand) networks at targeted node removal.
	}
	\label{fig:heat2D}    
\end{figure}

Proceeding to other thermodynamic functions, we rely mostly on some classical results \cite{Estrada&Hatano:2007,bianconi2009entropy,Estrada_etal:2012}.
Bianconi \cite{bianconi2009entropy} developed the statistical mechanics of network ensembles, defining entropy and free energy in direct analogy with physical systems, thereby establishing a rigorous thermodynamic interpretation of networks, while Estrada \textit{et al.} \cite{Estrada&Hatano:2007,Estrada_etal:2012} introduced the communicability framework and spectral partition functions for complex networks, using adjacency eigenvalues to define thermodynamic analogs such as entropy and free energy.
Following these approaches, we define the partition function
\begin{align}
\displaystyle Z(\beta) = \sum_i \re^{\beta\lambda_i},
\end{align}
where $\beta=1/T^*$ with $T^*$ being the formal thermodynamic temperature introduced as a control parameter, and heat capacity as 
\begin{align}\label{heat-def}
\displaystyle  C(\beta) = \beta^2\left[
\frac{Z''(\beta)}{Z(\beta)} - 
\left(\frac{Z'(\beta)}{Z(\beta)}\right)^2
\right].
\end{align}
Calculated heat capacity dependences on temperature and the number of removed nodes are shown in figures~\ref{fig:heat3D} and \ref{fig:heat2D}.

Heat capacity curves reveal one or two transition-like regimes as temperature varies. A single peak indicates one dominant spectral scale, while two peaks reflect a spectrum with multiple competing scales corresponding to distinct structural layers of the virus--host network.
Multiple maxima in heat-capacity curves were also observed in theoretical studies of frustrated magnetic systems and lattice models, where they similarly reflect the competing energy scales and hierarchical spectral features in the underlying Hamiltonian \cite{jurcisin2018multipeak,luo2021unusual,karlova2026}.

Now, we present some possible interpretations of heat capacity curve shapes.
The first peak (at low temperatures) denotes a competition among the largest eigenvalues of the adjacency spectrum; it thus highlights the role of the most dominant interaction modes.
The second peak (at higher temperatures) corresponds to the regime in which smaller eigenvalues start contributing significantly; it thus reflects additional structural scales or modular features of the network.
Such double peaks indicate the presence of two distinct spectral scales. The network structure is not characterized by a single dominant mode but by multiple layers of organization.
From the biological standpoint, the first peak may correspond to strongly connected core interactions (e.g., central viral or host biomolecules), while the second peak reflects contributions from more peripheral or context-specific interactions.

\section{Discussion}

The analysis of viral interactions shows that networks retain their structural integrity even after the removal of a considerable number of nodes. This indicates a high degree of robustness and resilience in such networks, which may complicate the development of effective strategies for combating viruses. The obtained results can serve as supplementary material for future studies of viral infections and the development of treatment methods based on microRNA regulation.

Considering the directed and weighted connections within the structure of viral networks enables a more detailed examination of interaction dynamics, enriching the analysis with additional characteristics. However, such an approach considerably complicates the interpretation of the results and necessitates the development of specific methods for data processing and visualization. The observed highly skewed distributions suggest the existence of a narrow group of critically important elements that play a central role in the network; this, in turn, may have important implications for modelling viral interaction mechanisms and identifying the strategies for their neutralization.

The irregular behavior of various parameters under different node removal approaches also highlights the need for a more detailed analysis of the local and global stability of the network. The identified strongly connected components indicate the presence of functionally significant clusters that may be promising targets for further research.

Heat capacity curves derived from adjacency spectra reveal a transition-like behavior: peaks mark structural scales at which the network reorganizes. For virus--host interactions, these might correspond to certain critical clusters that dominate the resilience.

Future directions include a detailed analysis of network motifs, which will allow for a deeper understanding of typical structural interaction patterns, as well as investigations into the influence of various factors on the resilience and the dynamics of viral networks.

\section*{Data Availability}
All data supporting the findings of this study have been deposited in Zenodo \cite{rovenchak_2026:DataSupplementary}.

\section*{Acknowledgement}
This work was partially supported by grant No. 0126U002265 from the Ministry of Education and Science of Ukraine.

\bibliographystyle{cmpj}
\bibliography{bibliography}

\begin{thebibliography}{10}
\providecommand{\url}[1]{\texttt{#1}}
\providecommand{\urlprefix}{URL }
\expandafter\ifx\csname urlstyle\endcsname\relax
  \providecommand{\doi}[1]{doi:\discretionary{}{}{}#1}\else
  \providecommand{\doi}{doi:\discretionary{}{}{}\begingroup
  \urlstyle{rm}\Url}\fi
\providecommand{\eprint}[2][]{\url{#2}}

\bibitem{fu2011imitation}
Fu~F., Rosenbloom~D.~I., Wang~L., Nowak~M.~A., Proc. R. Soc. B, 2011,
  \textbf{278}, No. 1702, 42--49, \doi{10.1098/rspb.2010.1107}.

\bibitem{damas2020broad}
Damas~J., Hughes~G.~M., Keough~K.~C., Painter~C.~A., Persky~N.~S., Corbo~M.,
  Hiller~M., Koepfli~K., Pfenning~A.~R., Zhao~H., Genereux~D.~P., Swofford~R.,
  Pollard~K.~S., Ryder~O.~A., Nweeia~M.~T., Lindblad-Toh~K., Teeling~E.~C.,
  Karlsson~E.~K., Lewin~H.~A., Proc. Natl. Acad. Sci. U.S.A., 2020,
  \textbf{117}, No.~36, 22311--22322, \doi{10.1073/pnas.2010146117}.

\bibitem{gulbahce2012viral}
Gulbahce~N., Yan~H., Dricot~A., Padi~M., Byrdsong~D., Franchi~R., Lee~D.-S.,
  Rozenblatt-Rosen~O., Mar~J.~C., Calderwood~M.~A., et~al., PLoS Comput. Biol.,
  2012, \textbf{8}, No.~6, e1002531, \doi{10.1371/journal.pcbi.1002531}.

\bibitem{vidal2011interactome}
Vidal~M., Cusick~M.~E., Barab{\'a}si~A.-L., Cell, 2011, \textbf{144}, No.~6,
  986--998, \doi{10.1016/j.cell.2011.02.016}.

\bibitem{guirimand2015virhostnet}
Guirimand~T., Delmotte~S., Navratil~V., Nucleic Acids Res., 2015, \textbf{43},
  No.~D1, D583--D587, \doi{10.1093/nar/gku1121}.

\bibitem{avs2022virus}
Krishna~S. A. V.~S., Sinha~S., Donakonda~S., Comput. Struct. Biotechnol. J.,
  2022, \textbf{20}, 4025--4039, \doi{10.1016/j.csbj.2022.07.040}.

\bibitem{bosl2019common}
B{\"o}sl~K., Ianevski~A., Than~T.~T., Andersen~P.~I., Kuivanen~S., Teppor~M.,
  Zusinaite~E., Dumpis~U., Vitkauskiene~A., Cox~R.~J., et~al., Front. Immunol.,
  2019, \textbf{10}, 2186, \doi{10.3389/fimmu.2019.02186}.

\bibitem{fendt2022overview}
Fendt~S.-M., Ralser~M., Curr. Opin. Syst. Biol., 2022, \textbf{31}, 100432,
  \doi{10.1016/j.coisb.2022.100432}.

\bibitem{lasso2019structure}
Lasso~G., Mayer~S.~V., Winkelmann~E.~R., Chu~T., Elliot~O.,
  Patino-Galindo~J.~A., Park~K., Rabadan~R., Honig~B., Shapira~S.~D., Cell,
  2019, \textbf{178}, No.~6, 1526--1541, \doi{10.1016/j.cell.2019.08.005}.

\bibitem{zhou2020network}
Zhou~Y., Hou~Y., Shen~J., Huang~Y., Martin~W., Cheng~F., Cell Discovery, 2020,
  \textbf{6}, No.~1, 14, \doi{10.1038/s41421-020-0153-3}.

\bibitem{sarkanych2024consensus}
Sarkanych~P., Sevinchan~{\relax Yu}., Krasnytska~M., Romanczuk~P.,
  Holovatch~{\relax Yu}., Condens. Matter Phys., 2024, \textbf{27}, No.~3,
  33801, \doi{10.5488/cmp.27.33801}.

\bibitem{sarkanych2016universality}
Sarkanych~P., Holovatch~{\relax Yu}., Kenna~R., Mac~Carron~P., J. Phys. Stud.,
  2016, \textbf{20}, No.~4, 4801, \doi{10.30970/jps.20.4801}.

\bibitem{holovatch2018statistical}
Holovatch~{\relax Yu}., Dudka~M., Blavatska~V., Palchykov~V., Krasnytska~M.,
  Mryglod~O., J. Phys. Stud., 2018, \textbf{22}, No.~2, 2801,
  \doi{10.30970/jps.22.2801}.

\bibitem{virbase3.0:www}
{\relax ViRBase v3.0}, {ViRBase v3.0: Virus-Host ncRNA Interaction Database},
  2021, [Online; accessed 22-Jun-2025],
  \urlprefix\url{https://www.rna-society.org/virbase/}.

\bibitem{li2015virbase}
Li~Y., Wang~C., Miao~Z., Bi~X., Wu~D., Jin~N., Wang~L., Wu~H., Qian~K., Li~C.,
  et~al., Nucleic Acids Res., 2015, \textbf{43}, No.~D1, D578--D582,
  \doi{10.1093/nar/gku903}.

\bibitem{cheng2022virbase}
Cheng~J., Lin~Y., Xu~L., Chen~K., Li~Q., Xu~K., Ning~L., Kang~J., Cui~T.,
  Huang~Y., Zhao~X., Wang~D., Li~Y., Su~X., Yang~B., Nucleic Acids Res., 2022,
  \textbf{50}, No.~D1, D928--D933, \doi{10.1093/nar/gkab1029}.

\bibitem{zipf2013psycho}
Zipf~G.~K., The Psycho-Biology of Language: An Introduction to Dynamic
  Philology, Routledge, 2013, \doi{10.4324/9781315009421}.

\bibitem{kalankesh2012language}
Kalankesh~L.~R., Stevens~R., Brass~A., BMC Bioinf., 2012, \textbf{13}, 127,
  \doi{10.1186/1471-2105-13-127}.

\bibitem{rovenchak2018telling}
Rovenchak~A., Mod. Phys. Lett. B, 2018, \textbf{32}, No.~05, 1850057,
  \doi{10.1142/S0217984918500574}.

\bibitem{semple2022linguistic}
Semple~S., {\relax Ferrer-i-Cancho}~R., Gustison~M.~L., Trends Ecol. Evol.,
  2022, \textbf{37}, No.~1, 53--66, \doi{10.1016/j.tree.2021.08.012}.

\bibitem{laherrere1998stretched}
Laherr{\`e}re~J., Sornette~D., Eur. Phys. J. B, 1998, \textbf{2}, No.~4,
  525--539, \doi{10.1007/s100510050276}.

\bibitem{rovenchak2018diary}
Rovenchak~A., Riley~C., Sherman~T., J. Quant. Linguist., 2017, \textbf{25},
  No.~3, 271--287, \doi{10.1080/09296174.2017.1373510}.

\bibitem{ferrer-i-cancho_sole2001}
{\relax Ferrer-i-Cancho}~R., Sol{\'e}~R.~V., J. Quant. Linguist., 2001,
  \textbf{8}, No.~3, 165--173, \doi{10.1076/jqul.8.3.165.4101}.

\bibitem{buk_rovenchak2004}
Buk~S.~N., Rovenchak~A.~A., J. Quant. Linguist., 2004, \textbf{11}, No.~3,
  161--171, \doi{10.1080/0929617042000314912}.

\bibitem{piantadosi2014}
Piantadosi~S.~T., Psychon. Bull. Rev., 2014, \textbf{21}, No.~5, 1112--1130,
  \doi{10.3758/s13423-014-0585-6}.

\bibitem{holovatch_palchykov2016}
Holovatch~{\relax Yu}., Palchykov~V., In: Maths Meets Myths: Quantitative
  Approaches to Ancient Narratives, Kenna~R., MacCarron~M., MacCarron~P.
  (Eds.), Springer International Publishing, Cham, 2016, 159--175,
  \doi{10.1007/978-3-319-39445-9_9}.

\bibitem{rovenchak_buk2018}
Rovenchak~A., Buk~S., J. Quant. Linguist., 2018, \textbf{25}, No.~1, 1--21,
  \doi{10.1080/09296174.2017.1324601}.

\bibitem{simon1955}
Simon~H.~A., Biometrika, 1955, \textbf{42}, No. 3/4, 425--440,
  \doi{10.2307/2333389}.

\bibitem{newman2005}
Newman~M. E.~J., Contemp. Phys., 2005, \textbf{46}, No.~5, 323--351,
  \doi{10.1080/00107510500052444}.

\bibitem{mirbase:www}
miRBase, {miRBase: the microRNA database}, [Online; accessed 22-Jun-2025],
  \urlprefix\url{https://mirbase.org/}.

\bibitem{Pettersen_etal:2004}
Pettersen~E.~F., Goddard~T.~D., Huang~C.~C., Couch~G.~S., Greenblatt~D.~M.,
  Meng~E.~C., Ferrin~T.~E., \mbox{J. Comput. Chem.}, 2004, \textbf{25}, No.~13,
  1605--1612, \doi{10.1002/jcc.20084}.

\bibitem{estrada2012structure}
Estrada~E., The Structure of Complex Networks: Theory and Applications, Oxford
  Academic Press, 2011, \doi{10.1093/acprof:oso/9780199591756.001.0001}.

\bibitem{bianconi2009entropy}
Bianconi~G., Phys. Rev. E, 2009, \textbf{79}, No.~3, 036114,
  \doi{10.1103/physreve.79.036114}.

\bibitem{callaway2000network}
Callaway~D.~S., Newman~M. E.~J., Strogatz~S.~H., Watts~D.~J., Phys. Rev. Lett.,
  2000, \textbf{85}, No.~25, 5468, \doi{10.1103/PhysRevLett.85.5468}.

\bibitem{albert2000error}
Albert~R., Jeong~H., Barab{\'a}si~A.-L., Nature, 2000, \textbf{406}, No. 6794,
  378--382, \doi{10.1038/35019019}.

\bibitem{barrat2008dynamical}
Barrat~A., Barthelemy~M., Vespignani~A., Dynamical Processes on Complex
  Networks, Cambridge University Press, 2008, \doi{10.1017/CBO9780511791383}.

\bibitem{castellano2009statistical}
Castellano~C., Fortunato~S., Loreto~V., Rev. Mod. Phys., 2009, \textbf{81},
  No.~2, 591--646, \doi{10.1103/RevModPhys.81.591}.

\bibitem{Estrada&Hatano:2007}
Estrada~E., Hatano~N., Chem. Phys. Lett., 2007, \textbf{439}, No. 1--3,
  247--251, \doi{10.1016/j.cplett.2007.03.098}.

\bibitem{Estrada_etal:2012}
Estrada~E., Hatano~N., Benzi~M., Phys. Rep., 2012, \textbf{514}, No.~3,
  89--119, \doi{10.1016/j.physrep.2012.01.006}.

\bibitem{jurcisin2018multipeak}
Jur\v{c}i\v{s}inov\'a~E., Jur\v{c}i\v{s}in~M., Phys. Rev. E, 2018, \textbf{97},
  No.~5, 052129, \doi{10.1103/PhysRevE.97.052129}.

\bibitem{luo2021unusual}
Luo~Q., Hu~S., Kee~H.-Y., Phys. Rev. Res., 2021, \textbf{3}, 033048,
  \doi{10.1103/PhysRevResearch.3.033048}.

\bibitem{karlova2026}
Karlova~K., Rufino~A., Verkholyak~T., Caci~N., Wessel~S., Stre{\v{c}}ka~J.,
  Mila~F., Honecker~A., \mbox{Preprint \arxiv{2601.14382}}, 2026.

\bibitem{rovenchak_2026:DataSupplementary}
Rovenchak~A., Husiev~M., {Data and supplementary materials for: Thermodynamic
  stability and structural transitions in virus--host networks}, Zenodo, 2026,
  \doi{10.5281/ZENODO.19770307}.

\end{thebibliography}

\ukrainianpart

\title{Термодинамічна стабільність та структурні переходи
в мережах вірус--господар}
\author{А. Ровенчак\refaddr{label1,label2}, М. Гусєв\refaddr{label1}}
\addresses{
\addr{label1}Кафедра теоретичної фізики імені професора Івана Вакарчука, Львівський національний університет імені Івана Франка,
вул. Драгоманова, 12, 79005 Львів, Україна
\addr{label2}SoftServe, Inc., вул. Садова, 2-д, 79021 Львів, Україна
}

\makeukrtitle

\begin{abstract}
\tolerance=3000%
Розуміння взаємодії вірус--господар має вирішальне значення для прогнозування стабільності мереж за різних збурень.
У цьому дослідженні ми представляємо аналіз мереж, пов'язаних з вірусами, для кількох організмів (\textsl{Homo sapiens}, \textsl{Mus musculus}, \textsl{Gallus gallus}), що охоплює спрямовані та зважені зв'язки.
Ми обчислюємо низку параметрів мережі, включаючи топологічні характеристики та термодинамічні
величини, отримані зі спектрів суміжності, щоб отримати уявлення про структурну стійкість та динамічну поведінку
мереж. Для оцінки стабільності ми моделюємо два різних сценарії видалення вузлів: цілеспрямоване видалення найбільш впливових вузлів та випадкове видалення. Наші результати показують перехідну поведінку в спектральних термодинамічних функціях та характерні зміни в структурних показниках, що сприяє оцінці потенціалу термодинамічної основи для вивчення мереж вірус--господар та глибшого розуміння їхньої динаміки.
\keywords складні мережі, термодинаміка, стабільність мережі--стійкість мережі, біомолекули--віруси

\end{abstract}

\end{document}